\documentclass{aa}
\usepackage{graphicx,lscape}
\usepackage{txfonts}

\usepackage[]{natbib}
\begin{document}

\title{Radio structure of the blazar 1156$+$295 with  sub-pc resolution}
   \author{W.~Zhao\inst{1,2}
         \and X.-Y.~Hong\inst{1}
         \and T.~An\inst{1}
         \and D.-R.~Jiang \inst{1}
         \and Jun-Hui~Zhao \inst{3}
         \and L.~I.~Gurvits \inst{4,5}
         \and J.~Yang \inst{4}
      }

   \offprints{W.~Zhao}

   \institute{Shanghai Astronomical Observatory, CAS, 80~Nandan~Road, 200030~Shanghai, P.R.~China
              \and Graduate University of the Chinese Academy of Sciences, 100049~Beijing, P.R.~China
              \and Harvard-Smithsonian Center for Astrophysics, 60 Garden Street, MS 78, Cambridge, MA 02138, USA
              \and Joint Institute for VLBI in Europe, Postbus 2, 7990 AA Dwingeloo The Netherlands
              \and Institute of Space and Astronautical Science, Japan Aerospace Exploration Agency,
                   3-1-1 Yoshinodai Chuo-ku, Sagamihara, Kanagawa 252-5210,
                   Japan}
   \date{Received ..., 2010; accepted ..., 2010}


\abstract
{}
{1156$+$295 is a flat-spectrum quasar that is loud at both radio and
$\gamma$-ray wavelengths. Previous observations of the source revealed a
radio morphology on pc to kpc scales consistent with a helical jet
model. We carried out VLBA observations at centimeter and
millimeter wavelengths to study the structure of the
innermost jet and understand the relation between the
helical structure and astrophysical processes in the central
engine.}
{The source 1156$+$295 was observed with the VLBA at 86, 43, and
15~GHz at four epochs from 10 May 2003 to 13 March 2005. The
observations were carried out in a full polarization mode. The
highest resolution of the observations is 0.08~mas ($\sim$0.5~pc)
at 86~GHz.}
{A core-jet structure with six jet components is identified in
1156$+$295. Three jet components are detected for the first time.
The apparent transverse velocities of the six jet components
derived from proper motion measurements range between
3.6~$c$ and 11.6~$c$, suggesting that highly relativistic jet
plasma moves in the direction close to the line of sight. The
overall jet has an oscillating morphology with multiple
curvatures on pc scales, which might be indicative of a helical
pattern. Models of a helical jet are discussed in the context of both
Kelvin-Helmholtz (K-H) instability and jet precession. The
K-H instability model is in closer agreement with the observed
data.}
{The overall radio structure on scales from sub-pc to kpc
appears to be fitted with a hydrodynamic model in the
fundamental helical mode of Kelvin-Helmholtz (K-H) instability.
This helical mode with an initial characteristic wavelength of
$\lambda_0$=0.2~pc is excited at the base of the jet on the scale
of 0.005~pc (or $10^3R_s$, the typical size of the broad line
region of a supermassive black hole of $4.3\times10^8M_{\odot}$).
A precessing jet model can also fit the observed jet structure on
scales of between 10~pc and 300~pc. However, with the precessing jet model, additional
astrophysical processes may be required to explain the bendings of the inner jet
structure (1 to 10~pc) and re-collimation of the large-scale jet
outflow ($>$300 pc).}

\authorrunning{W. Zhao}
\titlerunning{Sub-pc structure of blazar 1156$+$295}
\maketitle

\section{Introduction}
\label{Intro}
Radio-loud AGNs have been observed for more than three decades,
and tremendous progress has been made in understanding the underlying physical
processes with detailed modeling ({\it e.g.} \citealt{Urry1995}). However, many
fundamental and important problems remain uncertain.
For example, it is unclear how the jets are formed, accelerated, and
collimated, and what causes the broad range of observed morphologies
of jets. The detailed observations of structures, polarizations,
and proper motions of jets on parsec and sub-parsec scales will
provide critical insight into how the observed morphologies
of radio jets are related to the dynamic processes of radio outflows
and accretion disks around single, perhaps binary, supermassive
black holes. The millimeter VLBI appears to be a unique tool to
image the extremely compact region of radio sources
associated with AGNs at the resolution of sub-milliarcsecond,
providing a wealth of information about the detailed jet physics.

The blazar 1156$+$295 ($z=0.729$) is an archetypal radio-loud AGN
with a prominent jet. It is variable across the entire spectrum.
Strong $\gamma$ -ray flares have been detected by EGRET and
Fermi at energies higher than 100 MeV (\citealt{Tho1995, Muk1997, Har1999,
AAA2009}). In the optical domain, the source has been classified
as an optically violent variable (OVV) source and highly polarized
quasar (HPQ) (\citealt{Wil1983, Wil1992, Gla1983}). In the radio
domain, 1156$+$295 displays a significant variability on timescales
from years to days ({\it e.g.} see the monitoring program of the
University of Michigan Radio Astronomy
Observatory\footnote{http://www.astro.lsa.umich.edu/obs/radiotel/gif/1156\_295.gif.}
and the calibration program of the NRAO Very Large
Array\footnote{http://www.vla.nrao.edu/astro/calib/polar};
\citealt{Lov2003, Sav2008}).

On a wide range of linear scales from pc to kpc, the source has
a ``core-jet" structure. The jet is curved and consists of several
bright knots. On pc scales, the jet initially points to the north
and then bends to the northeast at a distance of 3-4~mas from the
core. At about 25~mas from the core, the jet then bends northwest
and aligns itself with the initial direction of the kpc-scale
jet. On the kpc scales, the jet has a wiggling morphology around
a mean position angle of -18$\degr$ and terminates at a hotspot
2~arcsec north of the core (\citealt{Hon2004}). On the basis of the high
apparent super-luminal velocities (up to 15 $h^{-1}c$, where $h$ is the
Hubble constant in units of 100~km~s$^{-1}$~Mpc$^{-1}$, \citealt{McH1990}),
VLBI jet components, and ultra-high brightness temperatures
($>10^{12}K$) of the core component (\citealt{Pin1997, Jor2001,
Hon2004}), the angle between the jet axis and the line of sight is
quite small. The complex bending jet can be attributed to the
projected shape of a 3-D helical jet (\citealt{Hon2004}), which
appears to be common in radio-loud AGNs (see \citealt {Con1993}).

To identify the relation between the wiggling jet and the
activity of the central engine on sub-pc scales, we conducted
4-epoch VLBI observations of 1156$+$295 at three frequencies of
86, 43, and 15~GHz (3~mm, 7~mm, and 2~cm, respectively). On the basis
of observed morphologies, structural variations, and polarization
properties, kinematic models based on different physical mechanisms are presented.
Cosmological parameters of $H_0=71$~km~s$^{-1}$~Mpc$^{-1}$,
$\Omega_{M}=0.27$, and $\Omega_{\Lambda}=0.73$ are used throughout
this paper. At the distance of 1156$+$295, 1~mas corresponds to
7.3~pc, and a proper motion of 1~mas~yr$^{-1}$ corresponds to
23.7~$c$ (\citealt {Wri2006}).

\section{Observations and data reduction}

\begin{table*}
\caption{Observation log}
\label{Obslog}
 \begin{tabular}{c|c|c|c|c|c}
     \hline  \hline
Obs. date  &    Epoch  &  Frequency  & VLBA                  & Band Width & On-source Time        \\
           &    (yr)   &  (GHz)  &   participated telescopes & (MHz)      & (Min) \\\hline
2003-05-10 &   2003.36 &  15.0   &  All 10                   &  64        & 122 \\
           &           &  43.0   &  9 of 10 $^a$             &  64        & 122 \\
           &           &  86.0   &  7 of 10 $^{b, c, d}$     &  64        & 121 \\ \hline
2003-07-24 &   2003.56 &  15.0   &  9 of 10 $^e$             &  64        & 122 \\
           &           &  43.0   &  9 of 10 $^e$             &  64        & 122 \\
           &           &  86.0   &  8 of 10 $^{b, e}$        &  64        & 121 \\\hline
2004-04-01 &   2004.25 &  15.0   &  All 10                   &  64        & 118    \\
           &           &  43.0   &  All 10                   &  64        & 118    \\
           &           &  86.0   &  8 of 10 $^{b, c}$        &  64        & 117    \\\hline
2005-03-13 &   2005.19 &  15.0   &  9 of 10 $^f$             &  64        & 118    \\
           &           &  43.0   &  8 of 10 $^{f, g}$        &  64        & 118    \\
           &           &  86.0   &  8 od 10 $^{c, f}$        &  64        & 117    \\\hline

\end{tabular}
\\
Notes:
$a$: FD data were corrupted at 43.0~GHz;
$b$: BR had no 3~mm receiver;
$c$: SC had no 3~mm receiver;
$d$: No fringes detected at 86.0~GHz on baseline of HN;
$e$: SC was out for maintenance;
$f$: HN was removed during the observing run due to bad weather;
$g$: SC data were flagged out because of their abnormal system
temperature.
\end{table*}

\label{Obs} The multi-frequency observations were carried out with
the Very Long Baseline Array (VLBA) on 10 May 2003 (2003.36), 24
July 2003 (2003.56), 1 April 2004 (2004.25), and 13 March 2005
(2005.19) at 86, 43, and 15~GHz. Detailed information about the
observations is given in Table \ref{Obslog}.

We used a snapshot observing mode (one scan of 4-5 minutes duration)
for our main target source and calibrators. To optimize the
$u$-$v$ coverage, the three observing frequencies of 86, 43, and
15~GHz were cycled in each of the observing runs. A pointing scan
was inserted after each of the frequency switching to 86~GHz. A
total observing time around 6 hours on 1156$+$295 was evenly
allocated to three observing frequencies at each epoch.

All observations were made in the dual circular-polarization mode.
At 15 and 43~GHz, 3C273, J1310$+$3220 and OJ~287 were included in
the observing runs as calibrators for corrections for R-L
delay-offset, polarization leakage, and the absolute position angle
of the electric vector (EVPA), respectively. At 86~GHz, it is
difficult to find a compact calibrator with a stable polarization
angle to use for EVPA calibration, hence we only study the
total intensity structure of 1156$+$295 from the 86~GHz data. The
raw data were recorded in the 2-bit VLBA format with a total
bandwidth of 64~MHz and correlated in Socorro, New Mexico, USA.

The data were reduced following the standard procedure of VLBA
data reduction using the NRAO AIPS (Astronomical Image Processing
System) software package (\citealt{Dia1995}). The visibility
amplitudes were calibrated using the gain curves and system
temperature monitored during the observations. The atmospheric
opacity correction was applied by using the WX table, which is an
extension file of data in a table format containing ground weather
information. Fringe fitting was made using a short scan on 3C~273
to calibrate the delays and phases. The solutions were then
interpolated to the data on all the programme sources. Global
fringe fitting was carried out to help correct for the residual
delays and fringe rates. At 15 and 43~GHz, 3C~273 and OJ~287 were
used to calibrate the complex bandpasses.

A series of polarization-specific calibrations were made for the
15 and 43~GHz data. After the amplitude calibration, variations in
the parallactic angles were determined and removed from the data.
The R-L delay difference was determined by running the VLBA
procedure VLBACPOL on the highly polarized source 3C~273. The
instrumental polarization parameters of the antenna feeds were
calibrated with the AIPS task LPCAL using scans on J1310$+$3220.
The corrections were written in an updated antenna (AN) table and
then applied to the data. The calibration of the absolute
polarization angle was made from observations of OJ~287 or
J1310$+$3220. At 15~GHz, we used OJ~287, which is compact and
stable enough to be an EVPA calibrator. To obtain the EVPA
correction, we compared its apparent EVPA with the
near-simultaneous measurements included in the MOJAVE/2cm Survey
database\footnote{http://www.physics.purdue.edu/MOJAVE/sourcepages/1156+295.shtml}.
At 43~GHz, both OJ~287 and J1310+3220 were used for similar
calibrations. Corrections were derived by comparing their apparent
EVPA with the VLA/VLBA
measurements\footnote{http://www.vla.nrao.edu/astro/calib/polar}
at 43~GHz on close epochs. After calibration, the data were
divided into single-source files and imported into the Caltech
VLBI Program DIFMAP (\citealt{She1994}) for self-calibration and
imaging. The fully calibrated data at 15 and 43~GHz were then
re-imported into AIPS to generate polarization images. I, Q, and U
images were produced separately with the AIPS task IMAGR using natural
weighting. The polarization angle and polarization intensity
images were then obtained by combining the Q and U images using the
AIPS task COMB.

At 86~GHz, an antenna-based edition of the data was first produced in
DIFMAP to remove visibilities related to extremely large amplitude
fluctuations likely due to bad weather or imperfect atmospheric
opacity corrections. The $uv$ data were then averaged using Miriad (\citealt{Sau1995})
over a time span of 1 minute. We calculated the scalar average in amplitude
because the 86-GHz visibility phase is significantly affected by fast
atmospheric path length changes. After that, the non-imaging
analysis was done in the $uv$-domain (see the Section
\ref{Non86}). Finally, the 86-GHz data were re-imported into
DIFMAP for self-calibration and imaging. The image at 86-GHz was
made with uniform weighting to achieve a higher angular resolution.
A few iterations of phase-only self-calibration were made until
the {\it rms} noise in the image did not decrease. Amplitude
self-calibration was not performed because we found that it would cause
an uncertainty as large as 20-30 percent in the flux density at
86~GHz.

\section{Results and analysis}
\label{Res}

\subsection{Structure of the source 1156$+$295}
\label{Str}
\subsubsection{Images at 15 and 43~GHz}
\label{Ima15&43} 
\begin{figure*}
\includegraphics[height=0.9\textheight]{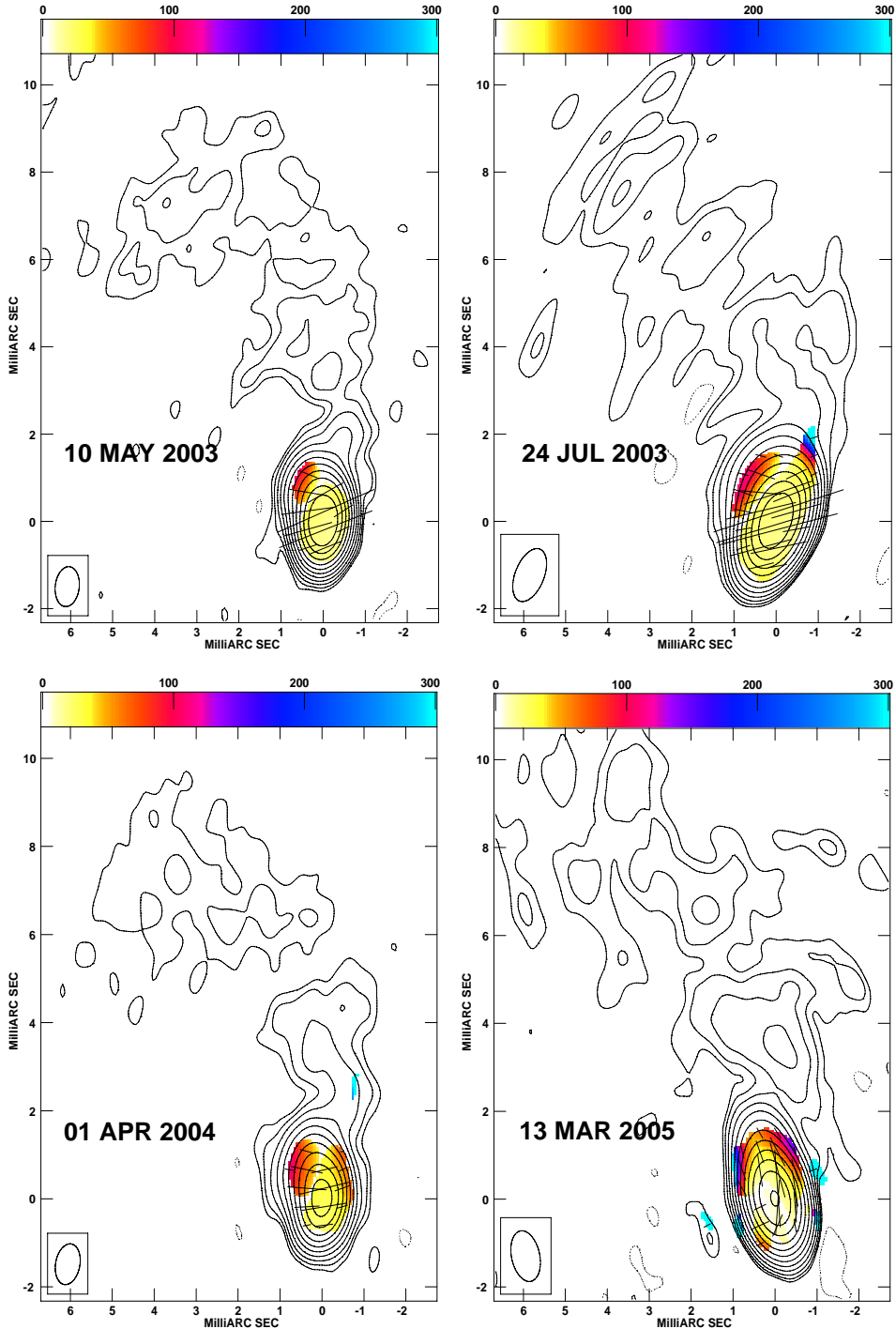}
\caption{The four epochs of VLBA images of 1156$+$295 at 15~GHz. The
contour maps represent Stokes I images. The pseudo-color maps
overlaid on contours represent the distribution of fractional
polarization in which linear polarization intensity above
4$\sigma$ is shown. The length of the short bars represents the
strength of the polarized emission,
1~mas$\sim$1.0$\times10^{-2}$~Jy~beam$^{-1}$. The orientation of
the short bars represent the apparent electric vector position
angle (EVPA) without a rotation measure correction. The detailed
parameters of the images are listed in Tables \ref{Par15&43} and
\ref{LP}.} \label{fig1}
\end{figure*}

The contour images in Figures \ref{fig1} and
\ref{fig2} represent the distribution of total radio intensity in
1156$+$295 at four epochs at 15 and 43~GHz, respectively. The
parameters of the images are listed in Table \ref{Par15&43}.

In the images at 15~GHz, the peak values of the intensity are 1.61, 1.86, 0.46, and 1.31~Jy~beam$^{-1}$
for the four epochs, respectively. At 43GHz, the values are 1.54, 2.22, 0.44, and 1.59~Jy~beam$^{-1}$, respectively.
The peak value at the highest epoch (2003.56) is four or five times as high as at the lowest epoch (2004.25).
This large variation in peak intensity is not due to calibration error, but reflects an intrinsic
variation in the source. The similar trend in the flux density variation is seen in the flux density
monitoring program at centimeter and millimeter wavelengths ({\it e.g.} VLA: 5, 8.4, 22, and 43~GHz, Mots$\ddot{a}$hovi: 22 and 37~GHz, OVRO MMA:110~GHz), confirming that the source started flaring around epoch 2003.3,
reached the peak around epoch 2003.6 and at higher frequencies earlier,
and that the total flux density decayed rapidly to its minimum value around epoch 2004.4.
The next flare appeared to reach its
peak around 2005.4.

In 15~GHz images, the source has a compact and bright region
(referred to as the "core" hereafter) that dominates the
brightness distribution and has a one-sided jet to the north.
The jet initially extends to the northeast, bends to the
northwest and becomes a diffuse structure at a distance of 3-4~mas
from the core. At about 12~mas from the core, the brightness of
the jet drops below the detection limit of
0.6~mJy~beam$^{-1}$(3$\sigma$). The overall structure of 1156$+$295
observed at 15~GHz is in good agreement with the previous
observations at lower frequencies (\citealt{Hon2004, Pin1997,
Jor2001}).


At 43~GHz, with an angular resolution of 0.2~mas
(corresponding to 1.5~pc, and close to the size of the broad
line region of a normal AGN), our observations reveal a core-jet
structure on a scale smaller than 1~mas (7.3~pc). The jet first extends
to the north in the innermost 0.5~mas, then bends to the
north-east along the direction at PA~$\sim$18$\degr$ with respect to
the core. The morphology is consistent with the structure observed
at 15~GHz.

\begin{table*}
\caption{The parameters of contour images at 15 and 43~GHz in
Figures \ref{fig1} and \ref{fig2}}
  \label{Par15&43}
\begin{tabular}{c|c|ccc|c|c|l}
     \hline\hline
     Frequency  & Epoch &       & Synthetized Beam       &               & $I_\mathrm{peak}$  & $1\sigma$       &Contours\\\cline{3-5}
                &         & Maj   &Min         &P.A.           &             &            &\\
    (GHz)       & (yr)    & (mas) &(mas)       &($\degr$)      &(Jy~beam$^{-1}$)    &(mJy~beam$^{-1}$)  &(mJy~beam$^{-1}$)\\\hline
     15.0       & 2003.36 & 0.95  &0.57        &-9.5           & 1.61        &0.20       &$0.60\times(-1,1,2,4,...,1024)$    \\
                & 2003.56 & 1.34  &0.74        &-21.7          & 1.86        &0.30       &$0.90\times(-1,1,2,4,...,1024)$    \\
                & 2004.25 & 0.94  &0.58        &-11.0          & 0.46        &0.20       &$0.60\times(-1,1,2,4,...,1024)$     \\
                & 2005.19 & 1.17  &0.67        &13.0           & 1.31        &0.20       &$0.60\times(-1,1,2,4,...,1024)$     \\\hline
     43.0       & 2003.36 & 0.35  &0.19        &-11.8          & 1.54        &0.40       &$1.20\times(-1,1,2,4,...,1024)$       \\
                & 2003.56 & 0.47  &0.27        &-23.2          & 2.22        &0.60       &$1.80\times(-1,1,2,4,...,1024)$         \\
                & 2004.25 & 0.33  &0.20        &6.2            & 0.44        &0.30       &$0.90\times(-1,1,2,4,...,256)$         \\
                & 2005.19 & 0.53  &0.34        &-10.9          & 1.59        &0.80       &$2.4\times(-1,1,2,4,...,512)$         \\\hline
\end{tabular}
  \end{table*}

\begin{table*}
\caption[]{15~GHz model-fitting results and physical parameters of
fitted models}
 \label{15model}
       \[
     \begin{tabular}{cccccccc|cccccc}
     \hline\hline
  Epoch    &Comp   &$S_\mathrm{c}$    &\textit{r}     &PA             &\textit{a}  & \textit{b / a}  &$\Phi$          &$T_\mathrm{b}$       & $\delta_\mathrm{eq}$ &$T_\mathrm{r}$               &$\gamma$ &$\theta$       \\
  (yr)     &       &(mJy)              &(mas)          &($\degr$)      &(mas)       &                 & ($\degr$)      & K                   &              & K                                   &         &($\degr$)       \\
 (1)      & (2)   &(3)                &(4)            &(5)            & (6)        & (7)             & (8)            &(9)                  &(10)          &(11)                                 &(12)     &(13)            \\\hline
   2001.17 & C     & 996.2 $\pm$149.4  & 0              & 0              & 0.21       & 0.16            & 43.4             & $2.0\times10^{12}$      &8.8           &$2.2\times10^{11}$                  &         &     \\
           & C5    & 115.3 $\pm$ 17.3  & 0.69$\pm$0.03  & 13 $\pm$3  & 0.35$\pm$0.03       & 1               & 0           & $1.3\times10^{10} $      &              &                                     &5.2      &4.5       \\
           & C4    & 6.2   $\pm$ 1.6   & 1.31$\pm$0.62  & -14$\pm$27 & 0.22$\pm$0.62       & 1               & 0           & $1.8\times10^{9} $      &              &                                     &5.5      &5.3      \\
           & C3    & 30.7  $\pm$ 32.0  & 4.32$\pm$ 1.91 & -8 $\pm$ 24 & 3.67$\pm$3.81      & 1               & 0           & $3.1\times10^{7} $      &              &                                     &--       &--     \\
           & C2    & 15.9  $\pm$ 14.7   & 6.69$\pm$0.89  & 18 $\pm$ 8 & 1.96$\pm$1.79      & 1               & 0           & $5.7\times10^{7} $      &              &                                     &--       &--     \\\hline
   2002.89 & C     & 2065.5$\pm$309.8& 0              & 0              & 0.37       & 0.31            & 4.7                & $0.7\times10^{12}$      &2.3           &$2.9\times10^{11}$                  &         & \\
           & C5    & 48.5  $\pm$  7.6& 0.69$\pm$0.11  & 23 $\pm$9  & 0.23$\pm$0.11       & 1               & 0             & $1.2\times10^{9} $      &              &                                     &4.2     &20.5      \\
           & C4    & 9.5   $\pm$ 2.6 & 1.86$\pm$0.58  & -16$\pm$18 & 0.52$\pm$0.58       & 1               & 0             & $4.8\times10^{8} $      &              &                                     &5.4     &19.1      \\
           & C3    & 35.9  $\pm$25.5   &4.51$\pm$0.87  & 9  $\pm$11   & 2.48$\pm$1.74       & 1               & 0             & $7.9\times10^{7} $      &              &                                     &--       &--      \\
           & C2    & 31.6  $\pm$25.8  &8.67$\pm$ 1.44 & 24 $\pm$ 10  & 3.57$\pm$2.88      & 1               & 0             & $3.4\times10^{7} $      &              &                                     &--       &--   \\\hline
   2003.36 & C     & 1605.5$\pm$240.8& 0              & 0              & 0.23       & 0.31            & 2.8                & $1.3\times10^{12}$      &5.4           &$2.5\times10^{11}$                  &         &     \\
           & C6    & 188.5 $\pm$ 28.3& 0.56$\pm$0.02  & 13 $\pm$2  & 0.33$\pm$0.02       & 1               & 0             & $2.3\times10^{10}$      &              &                                     &4.7      &10.1  \\
           & C5    & 41.3  $\pm$  6.2& 0.94$\pm$0.07  & 30 $\pm$4  & 0.33$\pm$0.07       & 1               & 0             & $5.1\times10^{9} $      &              &                                     &4.0      &9.5  \\
           & C4    & 9.4   $\pm$ 1.5 & 1.98$\pm$0.31  & -12$\pm$9  & 0.40$\pm$0.31       & 1               & 0             & $8.1\times10^{8} $      &              &                                     &4.4      &10.0  \\
           & C3    & 28.1  $\pm$  5.5& 4.56$\pm$0.22  & 1  $\pm$ 3 & 2.78$\pm$0.45       & 1               & 0             & $4.9\times10^{7} $      &              &                                     &--       &--  \\
           & C2    & 28.1  $\pm$ 14.5 &8.19$\pm$1.01  & 26 $\pm$ 7 & 3.93$\pm$2.02       & 1               & 0             & $2.5\times10^{7} $      &              &                                     &--       &--   \\\hline
   2003.56 & C     & 1907.5$\pm$286.1   & 0              & 0              & 0.28       & 0.31            & 15.7            & $1.1\times10^{12}$      &4.3           &$2.5\times10^{11}$                  &         & \\
           & C6    & 135.8 $\pm$ 20.4   & 0.59$\pm$0.04  & 17 $\pm$4  & 0.39$\pm$0.04       & 1               & 0          & $1.2\times10^{10} $      &              &                                     &4.6      &12.3   \\
           & C5    & 38.7  $\pm$  5.9   & 1.21$\pm$0.14  & 26 $\pm$7  & 0.40$\pm$0.14       & 1               & 0          & $3.2\times10^{9} $      &              &                                     &3.8      &12.5 \\
           & C4    & 29.0  $\pm$  4.5   & 2.25$\pm$0.18  & -10$\pm$5  & 0.94$\pm$0.18       & 1               & 0          & $4.5\times10^{8} $      &              &                                     &4.4      &12.8   \\
           & C3    & 28.2  $\pm$ 29.1    & 4.64$\pm$1.74  & 2  $\pm$21  & 3.38$\pm$3.49       & 1               & 0          & $3.3\times10^{7} $      &              &                                     &--       &--   \\
           & C2    & 24.3  $\pm$ 34.4    & 8.80$\pm$3.10  & 27 $\pm$19  & 4.38$\pm$6.19       & 1               & 0          & $1.7\times10^{8} $      &              &                                     & --      &--  \\\hline
   2004.11 & C     & 625.7 $\pm$ 93.9   & 0              & 0              & 0.31       & 0.23            & 9.1             & $0.4\times10^{12}$      &1.3           &$2.9\times10^{11}$                  &         & \\
           & C6    & 122.6 $\pm$ 18.4   & 0.57$\pm$0.02  & 19 $\pm$2  & 0.28$\pm$0.02       & 1               & 0          & $2.1\times10^{10}$      &              &                                     &8.8     &21.4 \\
           & C5    & 27.2  $\pm$  4.2   & 1.11$\pm$0.07  & 21 $\pm$4  & 0.41$\pm$0.07       & 1               & 0          & $2.2\times10^{9} $      &              &                                     &6.0     &24.7 \\
           & C4    & 11.5  $\pm$  3.1  &  1.96$\pm$0.17  & -12$\pm$5  & 1.33$\pm$0.17       & 1               & 0          & $8.8\times10^{7} $      &              &                                     &8.1     &22.1   \\
           & C3    & 21.8  $\pm$ 17.8    &4.73$\pm$1.10   & 2  $\pm$13  & 2.69$\pm$2.19       & 1               & 0          & $4.1\times10^{7} $      &              &                                     &--       &--  \\
           & C2    & 31.9  $\pm$ 19.1    &8.97$\pm$1.54   & 29 $\pm$10  & 5.17$\pm$3.09       & 1               & 0          & $1.6\times10^{6} $      &              &                                     &--       &--   \\\hline
   2004.25 & C     & 501.0 $\pm$ 75.2   & 0              & 0              & 0.43       & 0.34            & 9.5             & $0.1\times10^{12}$      &0.3           & --                                     &         &\\
           & C6    & 120.3 $\pm$ 18.1   & 0.62$\pm$0.01  & 18 $\pm$1  & 0.35$\pm$0.01       & 1               & 0          & $1.3\times10^{10} $      &              &                                     &--       &--   \\
           & C5    & 22.6  $\pm$  3.4   & 1.25$\pm$0.07  & 22 $\pm$3  & 0.50$\pm$0.07       & 1               & 0          & $1.2\times10^{9} $      &              &                                     &--       &--   \\
           & C4    & 6.3   $\pm$  1.1   & 2.33$\pm$0.23  & -18$\pm$6  & 0.60$\pm$0.23       & 1               & 0          & $2.4\times10^{8} $      &              &                                     &--       &--  \\
           & C3    & 26.0  $\pm$ 19.4   & 4.54$\pm$1.18  & 0  $\pm$ 15  & 3.17$\pm$2.36       & 1               & 0          & $3.5\times10^{7} $      &              &                                     &--       &--   \\
           & C2    & 25.3  $\pm$ 13.3    &8.94$\pm$1.44  & 32 $\pm$ 9   & 5.5 $\pm$2.88       & 1               & 0          & $1.1\times10^{7} $      &              &                                     &--       &--   \\\hline
   2005.19 & C     & 1307.9$\pm$196.2   & 0              & 0              & 0.08       & 1               & 0               & $2.5\times10^{12}$      &11.4          &$2.2\times10^{11}$                  &         & \\
           & C6    & 64.5  $\pm$   9.7  & 0.89$\pm$0.02  & 15 $\pm$1  & 0.33$\pm$0.02        & 1               & 0         & $7.9\times10^{9} $      &              &                                     &6.6      &3.4 \\
           & C5    & 11.9  $\pm$   1.8  & 1.27$\pm$0.09  & 20 $\pm$4  & 0.40$\pm$0.09        & 1               & 0         & $1.0\times10^{9} $      &              &                                     &6.3      &2.9   \\
           & C4    & 5.0   $\pm$   1.1  & 2.43$\pm$0.43  & -11$\pm$10 & 1.16$\pm$0.43        & 1               & 0         & $5.1\times10^{7} $      &              &                                     &6.6      &3.3  \\
           & C3    & 26.1  $\pm$ 8.5    & 4.09$\pm$0.33  &  4 $\pm$5  & 2.07$\pm$0.67         & 1               & 0         & $8.3\times10^{7} $      &              &                                     &--       &--   \\
           & C2    & 48.8  $\pm$ 11.3    &9.43$\pm$0.79  & 28 $\pm$5  & 6.94$\pm$1.58         & 1               & 0         & $1.4\times10^{7} $      &              &                                     &--       &--  \\\hline
\end{tabular}
   \]

Note: $S_{c}$ is the flux density of the fitted components;
\textit{r} and PA are the radial separation and the position angle
of the jet components with respect to the core component, respectively;
$a$ and $b$ are the major and minor axis size for the Gaussian
model in mas; $T_\mathrm{b}$ is the brightness temperature in the
parent-galaxy rest-frame, $\delta_\mathrm{eq}$ is the
equipartition Doppler factor and $T_\mathrm{r}$ is the intrinsic
brightness temperature eliminating the Doppler boosting; $\gamma$
and $\theta$  are the Lorentz factor and the viewing angle of jet
components.
  \end{table*}

\begin{table*}
\caption[]{43~GHz model-fitting results, and physical parameters
of fitted models}
 \label{43model}
       \[
     \begin{tabular}{cccccccc|ccccc}
     \hline\hline
  Epoch    &Comp   &$S_\mathrm{c}$    &\textit{r}     &PA             &\textit{a}  & \textit{b / a}  &$\Phi$          &$T_\mathrm{b}$       & $\delta_\mathrm{eq}$ &$T_\mathrm{r}$               &$\gamma$ &$\theta$       \\
   (yr)    &       &(mJy)              &(mas)          &($\degr$)      &(mas)       &                 & ($\degr$)      & K                   &              & K                                   &         &($\degr$)       \\
 (1)      & (2)   &(3)                &(4)            &(5)            & (6)        & (7)             & (8)            &(9)                  &(10)          &(11)                                 &(12)     &(13)            \\\hline
  2003.36  & C     & 1539.5$\pm$230.9    & 0          & 0             & 0.05       & 0.15            & -30.3          &$ 6.8\times10^{12}$ &13.4         &$5.1\times10^{11}$   &           &                              \\
           & C7    & 75.5  $\pm$ 11.4    & 0.22       & 15          & 0.06       & 1.00            & 0                &$ 3.5\times10^{10}$ &              &                        &7.3        &2.2                          \\
           & C6    & 105.3 $\pm$ 15.8    & 0.46       & 4           & 0.30       & 1.00            & 0                &$ 1.9\times10^{9}$  &              &                        &--        &--                          \\
           & C5    & 41.4  $\pm$  6.3    & 0.80       & 23          & 0.28       & 1.00            & 0                &$ 8.8\times10^{8}$  &              &                        &--        &--                           \\\hline
  2003.56  & C     & 2177.5$\pm$326.6    & 0          & 0             & 0.05       & 0.42            & -34.2          &$ 3.4\times10^{12}$ &16.5         &$2.1\times10^{11}$   &           &                            \\
           & C7    & 172.4 $\pm$ 25.9    & 0.29       & 21         & 0.16       & 1.00            & 0                 &$ 1.1\times10^{10}$  &              &                        &8.7        &1.5                         \\
           & C6    & 62.9  $\pm$  9.6    & 0.63       & 14          & 0.44       & 1.00            & 0                &$ 5.4\times10^{8}$  &              &                        &--        &--                        \\
           & C5    & 9.9   $\pm$  2.1    & 1.11$\pm$0.03 & 26 $\pm$2   & 0.41$\pm$0.03       & 1.00            & 0    &$ 2.9\times10^{7}$  &              &                        &--        &--                        \\\hline
  2004.25  & C     & 482.6 $\pm$72.4     & 0          & 0           & 0.10       & 0.40            & -1.2             &$ 0.2\times10^{12}$ &0.6          & --   &           &                          \\
           & C7    & 38.4  $\pm$ 5.8     & 0.38       & -3      & 0.12       & 1.00            & 0                    &$ 4.4\times10^{9}$  &              &                        &  --           &   --                     \\
           & C6    & 61.2  $\pm$ 9.2     & 0.62       & 18     & 0.23       & 1.00            & 0                     &$ 1.9\times10^{9}$  &              &                        &  --           &   --                     \\
           & C5    & 27.5  $\pm$ 4.3     & 1.17$\pm$0.01& 16$\pm$1      & 0.40$\pm$0.01       & 1.00            & 0   &$ 2.9\times10^{8}$  &              &                        &   --          &   --                    \\\hline
  2005.19  & C     & 1648.3$\pm$247.3    & 0            & 0            & 0.10       & 0.60            & -89.0         &$ 0.5\times10^{12}$ &6.8          &$0.7\times10^{11}$   &           &                          \\
           & C6    & 24.4  $\pm$  4.1    & 0.78$\pm$0.02       & 12$\pm$2     & 0.27$\pm$0.02       & 1.00     & 0    &$ 5.5\times10^{8}$  &              &                        &--   & --                   \\\hline
\end{tabular}
   \]
Note: $S_{c}$ is the flux density of the fitted components;
\textit{r} and PA are the radial separation and the position angle
of the jet components with respect to the core component respectively;
$a$ and $b$ are the major and minor axis size for the Gaussian
model in mas; $T_\mathrm{b}$ is the brightness temperature in the
parent-galaxy rest-frame, $\delta_\mathrm{eq}$ is the
equipartition Doppler factor and $T_\mathrm{r}$ is the intrinsic
brightness temperature eliminating the Doppler boosting; $\gamma$
and $\theta$  are the Lorentz factor and the viewing angle of jet
components.
  \end{table*}

\subsubsection{Model fitting of data at 15 and 43GHz}
\label{Mod15&43}

To quantitatively analyze the properties of 1156$+$295,
we fitted the source emission structure with Gaussian components
using the program MODELFIT in DIFMAP. The model-fitting parameters
are listed in the left panels of Tables \ref{15model} and \ref{43model}
for 15 and 43~GHz, respectively.

At 15~GHz, in addition to four sets of 15~GHz data described in
Section \ref{Obs}, three additional data sets for epochs 2001.17,
2002.90, and 2004.11 from the "MOJAVE 2cm Survey Data Archive"  are
used in the model fitting. The source structure is most closely
fitted with six Gaussian components (see the left panel of Figure
\ref{fig4}) at this frequency. The brightest region (component C)
could be fitted with the elliptical Gaussian component that was
considered to be the core in the previous works
(\citealt{Hon2004}). The jet is fitted with five circular Gaussian
components, which are named `C2' to `C6' following the
nomenclature in \citet{Hon2004}. The component C5 was first detected at the
epoch 2001.17 with a flux density of 115.3~mJy at a distance of
$\sim$0.7~mas in PA$\sim$13$\degr$. Its flux density dropped
quickly to about 10$\%$ of its initial value at the epoch 2005.19
and the size of the component expanded slightly as C5 moved out from the core.
C6 was first detected at the epoch 2003.36 with a flux density
of 188.5~mJy. In comparison to C5, both the size and flux density
of C6 varied relatively slowly as it moved out from the
core.

At 43~GHz, in the first three epochs (2003.36, 2003.56, and
2004.25), both C5 and C6 were detected. In addition, C7, the
closest component to the core was detected (see the right panel of
Figure \ref{fig4}). All the components C5, C6, and C7 were
detected for the first time at the epoch of 2003.36 at this
frequency. We found C6 varied significantly and the variation in
the flux density at 43~GHz appeared to be correlated with that
observed at 15~GHz. We also note that from 2003.36 to
2003.56, the apparent flux density of C7 increased by a factor of 2.3 and
the apparent size increased by a factor of 2.7. The possible reason for the
rapid increase in the radiation intensity of C7
is the absence of the SC (San Croix) antenna which decreases the resolution and causes an increase in the
apparent size, hence the apparent flux density of C7.
In the fourth epoch (2005.19), only two components
were detected because of the relatively poorer sensitivity and angular resolution in the absence of two VLBA antennas (HN and
SC); the data were fitted with two Gaussian components.

\subsubsection{The visibility analysis and images of 86GHz data}
\label{Non86}

We inspected the visibility amplitude of 86~GHz data
as a function of the projected baseline to acquire a general idea
of the source structure. The visibility amplitude profiles are not
in consistent with the expectations for a simple point source, but there is some evidence of intrinsic source structures.

We fitted the amplitude profiles on three different baseline
length ranges, the shortest (60-70$\times10^{6}\lambda$),
intermediate (100-200$\times10^{6}\lambda$), and longest
(1100-1500$\times10^{6}\lambda$), with a point source model,
respectively. The correlated flux density on the shortest
baselines ($S_{s}$) is 1.5, 1.6, 0.8, and 0.7~Jy at epoch
2003.36, 2003.56, 2004.25, and 2005.19, respectively (see Table
\ref{86model}), suggesting a flux density variation on a timescale
 of years to months (the shortest time separation being 2.4
months). The largest variation accounts for a factor of 2.3.
The trend in the flux density variation observed at 86~GHz agrees with that seen
for the 15 and 43~GHz observations at these epochs, and the variations in flux density reported from the
monitoring program at centimeter and millimeter wavelengths (see Section \ref{Ima15&43}).
The correlated flux densities on the intermediate baseline ($S_m$) and longest
baseline ($S_l$) are lower than $S_s$, providing evidence that
the source structure at 86~GHz is resolved by the VLBA at these
epochs.

We then use the brightness distribution models introduced by
\citealt{Pea1995}) to estimate the flux density ($S_{f}$) and
size ($\theta_{d}$) of the source. For the first three epochs, an
extended source model can be used (see Figure \ref{86GHzNon}). The
calculation gives the flux density 1.7, 1.8, and 0.8~Jy,
respectively, and the largest size of the source 0.06~mas
(2003.36), 0.05~mas (2003.56), and 0.05~mas (2004.25), respectively
(see Table \ref{86model}). For the first three epochs, the
correlated flux density on the shortest baseline ($S_s$) is
comparable to the fitted Gaussian peak of the amplitude profile,
implying that $S_s$ is a reasonable representative of the
zero-spacing flux density. For the last epoch (2005.19), a double-point source model was used (see Figure \ref{86GHzNon}). The flux
density is 0.9~Jy, the flux density ratio of two point sources is 3:1,
and the separation of the two point components is 0.16~mas (see Table
\ref{86model}).

On 86~GHz CLEANed images (see Figure \ref{fig3}), most of the
jet emission is resolved out at an angular resolution of
0.2~mas $\times$0.08~mas. At the epoch 2003.36, a weak jet is seen
$\sim$0.3~mas north of the core. The 86~GHz jet coincides with
the innermost jet component C7 detected at 43~GHz at the same
epoch. Earlier work, based on Global Millimeter VLBI Array
observations on 26 October 2001 at 86~GHz with an angular
resolution of 0.188$\times$0.038~mas, detected a jet component
at 0.19~mas north of the core, marginally coinciding with
C7 in our observations. At all other three epochs of 2003.56,
2004.25, and 2005.19, the jet emission on scales of
between 0.1 and 1~mas detected at both 43 and 15~GHz is
below the detection threshold at 86~GHz (i.e., $3\sigma\simeq
30mJy~beam^{-1}$), showing only a point-like source in 1156$+$295.

\begin{table}
\caption{86~GHz parameters from model fitting of amplitude
profiles} \label{86model}
 \[
 \begin{tabular}{c|ccc|cc|c}
     \hline\hline
  Epoch    & $S_{s}$ &$S_{m}$   & $S_{l}$  & $S_{f}$     & $\theta_{d}$ & $T_\mathrm{b}$                         \\
   (yr)    & (Jy)    &(Jy)      & (Jy)     & (Jy)        & (Jy)         & K                       \\
   (1)     & (2)     &(3)       & (4)      & (5)         & (6)          & (7)                \\\hline
  2003.36  & 1.5    & 1.1       & 0.6      & 1.7        & 0.06         & $ 1.9\times10^{11}$             \\
  2003.56  & 1.6    & 1.2         & 1.3        & 1.8        & 0.05         & $ 2.9\times10^{11}$               \\
  2004.25  & 0.8    & 0.4         & 0.3        & 0.8        & 0.05         & $ 1.2\times10^{11}$                 \\
  2005.19  & 0.7    & 0.5       & 0.7      & 0.9$^{a}$  & 0.16         & --               \\\hline
   $Errors$&\multicolumn{5}{|c|}{20$\%$}                                       &  --             \\\hline
\end{tabular}
\]
Note: $^a$ it is fitted with a double point-source model, the
separation
of the two points is 0.16~mas, the flux density ratio is 3:1.\\
$S_{s}$, $S_{m}$, and $S_{l}$ are the correlated flux density on
the shortest (60-70$\times10^{6}\lambda$), intermediate
(100-200$\times10^{6}\lambda$), and longest
(1100-1500$\times10^{6}\lambda$) baselines, respectively; $S_{f}$
and $\theta_{d}$ are the flux density and the size of the fitted
Gaussian model; $T_\mathrm{b}$ is the brightness temperature in
the parent-galaxy rest-frame.
\end{table}

\subsection{Flux variability of the core component}
\label{Flu}

The radio emission of the source is dominated by the compact core
at all three observing frequencies of our experiment. The
variability of the core component C at each frequency can be
assessed using the index of variability $V$

\begin{equation}
V=\frac{S_{max}-S_{min}}{S_{max}+S_{min}}
\end{equation}
where $S_{max}$ and $S_{min}$ are the maximum and minimum fitted flux densities of the core component.
A value of $V\sim 0$ represents no variability and $V\sim 1$ extreme
variability. During the observing period, $V$ was
0.58$\pm$0.07, 0.64$\pm$0.06, and 0.39$\pm$0.13 for 15, 43, and
86~GHz, respectively, suggesting that the component C is highly
variable (the uncertainties in $V$ are evaluated based on the uncertainties
in the $S_{max}$ and $S_{min}$).

The core of an AGN is often defined to be the most compact region of
the jet or the jet base, which always appears to be optically thick at
centimeter wavelengths because of the superposition of
sub-sections of the inner jet with various synchrotron
self-absorption turnover frequencies. At millimeter wavelengths,
many quasars and AGNs display a spectral turnover at
$10^{11\pm0.5}$~GHz (32-320~GHz)(\citealt{Mar1995}). In this
frequency range, the inner jet becomes optically thin, allowing us to observe
the intrinsic source structure with VLBI at mm-submm wavelengths. Figure \ref{fig4b} shows the flux density of the core
at three frequencies at each of four epochs. It shows that at the
epochs 2003.56 and 2005.19, the flux density at 43~GHz dominates
over those at 15 and 86~GHz, implying that the turnover frequency is around 43~GHz.

\begin{table*}
\caption{The parameters of contour images in Figure \ref{fig3}}
\label{Par86}
\[
\begin{tabular}{c|ccc|c|c|c|l}
     \hline\hline
    Epoch   &       & Synthetical Beam   &          & $I_\mathrm{peak}$  & $1\sigma$  &Contours                         & $S_{C}$     \\\cline{2-4}
            & Maj   &Min                 &P.A.      &                    &            &                                 &         \\
    (yr)    & (mas) &(mas)               &($\degr$) &(Jy~beam$^{-1}$)           &(mJy~beam$^{-1}$)  &(mJy~beam$^{-1}$)                       &  (Jy)         \\\hline
    2003.36 & 0.22  &0.08                &-24.3     & 0.92               & 4.99       &$15.0\times(-1,1,2,4,...,  32)$  & 1.27            \\
    2003.56 & 0.23  &0.08                &-24.7     & 1.96               & 10.00      &$30.0\times(-1,1,2,4,...,  64)$  & 2.02              \\
    2004.25 & 0.21  &0.08                &-25.1     & 0.58               & 4.20       &$12.6\times(-1,1,2,4,8,16)   $   & 0.58             \\
    2005.19 & 0.24  &0.11                &-11.4     & 0.73               & 5.90       &$17.7\times(-1,1,2,4,...,  64)$  & 0.74            \\\hline
\end{tabular}
\]
Note: $S_{C}$ is the CLEANed flux density
  \end{table*}

\begin{table*}
\caption{Linear polarization properties at 15 and 43~GHz}
\label{LP}
\[
\begin{tabular}{c|c|c|cccc}
     \hline\hline
     Frequency  & Epoch   & Component  & $F_\mathrm{peak}$   &$F_\mathrm{P}$          &$m$    &EVPA\\
    (GHz)       & (yr)    &       &(mJy~beam$^{-1}$)  &(mJy)          &($\%$) &($^{\circ}$)  \\\hline
     15.0       & 2003.36 &C      &28.9$\pm$0.3  &30.4$\pm$0.7     & 2.0$\pm$0.4   &112$\pm$ 7\\
                &         &C5     & 9.9$\pm$0.3  & 3.7$\pm$0.2     & 6.0 $\pm$2.2  & 75$\pm$ 6\\\cline{2-7}
                & 2003.56 &C      &34.6$\pm$0.2  &36.7 $\pm$0.8    & 2.2 $\pm$0.7  &106$\pm$ 7 \\
                &         &C5     & 9.6$\pm$0.2  & 4.4$\pm$0.2     &6.7$\pm$2.6   & 74$\pm$ 4 \\\cline{2-7}
                & 2004.25 &C      &12.8$\pm$0.1  &14.3 $\pm$0.3    & 3.6$\pm$1.1   &106$\pm$ 2 \\
                &         &C5     &10.6$\pm$0.1  & 5.5 $\pm$0.2    &6.7$\pm$2.0   & 80$\pm$ 3\\\cline{2-7}
                & 2005.19 &--     &18.2$\pm$0.3  &20.4 $\pm$0.2    & 4.8$\pm$4.4   &44$\pm$3\\\hline
     43.0       & 2003.36 &--     &36.6$\pm$0.5  &27 $\pm$0.8      & 2.4 $\pm$0.3  &120$\pm$ 8\\\cline{2-7}
                & 2003.56 &--     &29.1$\pm$1.0  &17.5 $\pm$1.0    & 2.7 $\pm$0.3  & 35$\pm$12 \\\cline{2-7}
                & 2004.25 &--     & 5.2$\pm$0.4  & 1.5 $\pm$0.2    & 1.8$\pm$0.5   & 72$\pm$ 5 \\\hline
\end{tabular}
\]
Note:\\$F_\mathrm{peak}$: Peak polarized brightness.\\
$F_\mathrm{P}$: Integrated polarized flux density.\\
$m$: the polarization percentage.\\
EVPA: Electric-Vector Position Angle.\\
\end{table*}

\subsection{Superluminal motion of jet components}
\label{proper}
The angular distances to the core of all detected jet components
were measured at all observing epochs. We
carried out linear regression fitting of the separation of the jet
components from the core as a function of the observing epoch.
From the fitted slope of the linear regression, we determined the
apparent velocities of each jet component.

The uncertainty in the separation of jet components to the
core component C is estimated to be $\Delta R=\frac{W}{2SNR}$ (for the compact components from C4 to C7),
or $\Delta R=\frac{a(1+SNR)^{1/2}}{2SNR}$ (for extended components C2 and C3), where
$W$ is the FWHM of the synthesized beam, $a$ is the major-axis size of the component, and $SNR$ is the
signal-to-noise ratio for a given component (\citealt{Foma2004, Lee2008}).

To better determine the proper motions of the components
C2, C3, and C4, in addition to the data discussed in the present
paper, we also included published data at lower observing
frequencies at the early epochs (\citealt{Hon2004}). For
components C5 and C6, both 15 and 43~GHz data from this work were
used in the analysis. Since C7 can be separated from the core
emission at only 43~GHz, the determination of the proper motion
for C7 involves only the 43~GHz data from this paper.

We use the inverse square of the uncertainty, (1/$(\Delta r)^2$),
as a weight parameter in the least squares fitting for C2, C3, C4,
and C7. For C5 and C6, we used both 15 and 43~GHz data points in
the least squares fitting. The inverse square of the difference
between the fitted positions at 15 and 43~GHz was used as a weight
parameter.

Figures \ref{fig51} and \ref{fig61} illustrate the linear fit of
the angular separation of the jet components for the determination
of their proper motions. Significant proper motions of 0.49$\pm$0.03,
$0.18\pm$0.03, 0.15$\pm$0.02, 0.19$\pm$0.02, and
0.16$\pm$0.04~mas~yr$^{-1}$ were determined for the jet components
C2, C4, C5, C6, and C7, respectively. The apparent transverse
velocities of 11.6$\pm$0.7~$c$, 4.3$\pm$0.7~$c$, 3.6$\pm$0.5~$c$,
4.5$\pm$0.5~$c$, and 3.8$\pm$0.9~$c$ derived from the proper
motions for C2, C4, C5, C6, and C7, respectively, confirming early
results on apparent superluminal motion in the blazar 1156$+$295
(\citealt{Hon2004}). We note that C3 shows a peculiar pattern in
its proper motion. The fitting to the data before 2001 gives a
proper motion of 0.41$\pm$0.08 mas~yr$^{-1}$ (or 9.7$\pm$1.9$c$).
No significant proper motions was detected after 2001.17 when the
component reached the place where the jet changes its direction by
$\sim40\degr$. Both the proper motion velocities and the component
sizes of the jet components show an increase beyond a distance
$>$2 mas (or 14~pc) away from the core. A possible explanation of these
changes is a decrease in the external pressure gradient.

We also note that the inferred values of apparent speed for C4, C5, C6, and C7 are similar,
and the apparent speeds of C2 and C3 are much higher than those of C4 to C7.
A possible explanation is that the extended components C2 and C3 are rather diffuse,
and the surface brightness of these components is very low. The uncertainty in the
radial distance from the core is large and might affect the fitting value of the apparent speed.

\subsection{Physical properties of compact components in VLBI images}
\label{Phy} The brightness temperature in the
parent-galaxy rest-frame $T_\mathrm{b}$ can be derived from the
measurements of the flux density ($S_{\rm ob}$), the observing
frequency ($\nu_{\rm ob}$), and the angular size ($\theta_{\rm
d}$) after correction for the cosmological expansion effect
$(1+z)$ (\citealt{AGRA1996})

\begin{equation}
T_\mathrm{b}= 1.77\times10^{12} (1+z) \frac{S_{\rm ob}}{\nu_{\rm
ob}^{2}\theta_{d}^{2}}\texttt{K},
\end{equation}
where $S_{\rm ob}$ is in Jy, $\nu_{\rm ob}$ in GHz, and $\theta_{d}$
is in mas. In our paper, $\theta_{d} =\sqrt{ab}$ for a Gaussian
component with major ($a$) and minor ($b$) axis sizes at 15 and
43~GHz.

The derived results of $T_\mathrm{b}$ are summarized in the Col.
9 of Tables \ref{15model} and \ref{43model}, and Col. 7 of Table
\ref{86model}. For some epochs at 15 and
43~GHz, we note that the observed brightness temperature $T_\mathrm{b}$ of the
core component is close to or exceeds the upper limit of
$T_\mathrm{b}$ restricted by the "inverse Compton
catastrophe"(\citealt{Kell1969}) or equipartition limit
(\citealt{Rea1994}). The higher brightness temperature
$\gtrsim10^{12}$K in flat-spectrum radio sources is often
considered to be a result of the Doppler boosting effect and /or
indications on a non-stationary process.

We note that at the epochs 2003.36 and 2003.56, the brightness
temperatures at 86~GHz are much lower than that measured at
43~GHz. This phenomenon seems to be common in flat-spectrum
radio-loud quasars and AGNs (\citealt{Lee2008, Lob2000}). Since
1156$+$295 is still rather compact at 86~GHz, {\it e.g.} the ratio
of the CLEANed flux density to the correlated flux density
is approximately equal to 1 in our observations. Therefore,
the relatively lower $T_\mathrm{b}$ at 86~GHz is not caused by the
loss of compactness (\citealt{Lee2008}) but may be caused by the intrinsic changes in the physical conditions
of the innermost jet ($<$0.04~mas, or 0.30~pc), {\it e.g.} the emission structure becomes optically thin
at 86~GHz and the brightness temperature is intrinsically lower than the
optically thick case.

To estimate the Doppler factor of 1156$+$295 using the results of our
observations, we assume the equipartition of the energy between
the radiating particles and the magnetic field. The equipartition
Doppler factor of a uniform self-absorbed source can be obtained
from Eq. 4 in \citet{AGRA1996} if we assume that the emission
arise from an optical-thin, homogeneous sphere.

$\delta_{\rm eq}=[[10^3
F(\alpha)]^{34}\{[1-(1+z)^{-1/2}]/(2h)\}^{-2} S^{16}_{\rm
op}\theta^{-34}_d\nonumber \times (1+z)^{15-2\alpha}(\nu_{\rm
op}\times 10^3)^{-(2\alpha+35)}]^{1/(13-2\alpha)}$ \\
where $h$ is the Hubble constant in units of 100~km~s$^{-1}$~Mpc$^{-1}$, the spectral index $\alpha$ is assumed to be -0.75,
and $\theta_{d}$ is the angular diameter of the homogenous sphere in mas. For elliptical
Gaussians, $\theta_{d}=1.8\,\sqrt{ab}$, where $a$ is the major axis size and $b$ is the minor axis size.
In addition, $\nu_{\rm op}$ is the turnover frequency in GHz.
To simplify the assessment, we used the observing frequency 15 and 43~GHz for replacing the turnover
frequency. The data at 86~GHz are not used here because the source might be
transparent to the synchrotron emission at this frequency. Thus, a
higher limit to the equipartition Doppler factor is inferred from
our estimate, where $S_{\rm op}$ is the flux density in Jy at the turnover frequency.
In this paper, we use the observed flux density at
15 and 43~GHz as the $S_{\rm op}$. $F(\alpha)$ is a factor with a value approximately equals to 3.4 at the turnover frequency.

The results are listed in column 10 of Tables \ref{15model} and \ref{43model}. The values
of the inferred Doppler factor are in the range of 1.3 to 16.5 except for the epoch 2004.25 when
the source is in the lowest state.

Since $\delta_\mathrm{eq}$ strongly depends on the observed values of
$S_{\rm op}$, $\nu_{\rm op}$, and $\theta_{d}$, the difference
between the values of $\delta_\mathrm{eq}$ at different
frequencies for a given epoch are dominated by the error in the
observed quantities. The intrinsic brightness temperature
eliminating the Doppler boosting in the source rest-frame is
$T_\mathrm{r}$, and $T_\mathrm{r}=T_\mathrm{b}/\delta$. Assuming
$\delta=\delta_\mathrm{eq}$, we can estimate the value of
$T_\mathrm{r}$.

The inferred physical quantities of $T_\mathrm{b}$,
$\delta_\mathrm{eq}$, and $T_\mathrm{r}$ are summarized in the right
panels of Tables \ref{15model}, \ref{43model}, and \ref{86model}.

The kinematic properties (the Lorentz factor $\gamma$, the viewing
angle $\theta$, and the intrinsic velocity $\beta$ in the unit of
$c$) for the compact components C4 to C7 (the components C2 and C3
appear too extended to have meaningful assessments for the
kinematic properties) in the inner jet region are discussed on the
basis of Equations (3--5) given in \citet{Ghi1993}

\begin{equation}
\gamma= \frac{\beta_{\rm app}^{2}+\delta^{2}+1}{2\delta},
\end{equation}

\begin{equation}
\theta=\arctan~(\frac{2\beta_{\rm app}}{\beta_{\rm
app}^{2}+\delta^{2}-1}),
\end{equation}

\begin{equation}
\beta=\frac{\beta_{\rm app}}{\sin\theta+\beta_{\rm
app}\cos\theta},
\end{equation}
where $\beta_{\rm_{app}}$ is the apparent proper motion velocity
of jet components derived by linear fitting in
Section \ref{proper} and $\delta$ is the Doppler factor with a
value derived from the equipartition calculation $\delta_\mathrm{eq}$.

The inferred Lorentz factor ranges between 3.8 and 8.8, the
viewing angle ranges between 1.5$\degr$ and 24.7$\degr$ (see Table
\ref{15model} and \ref{43model}), and the intrinsic jet speed
$\beta$ approximately equals to 0.98~$c$. The jet properties
inferred from the kinematic quantities confirm that 1156$+$295 is
a source affected by the extreme relativistic-beaming effect.

\subsection{Linear polarization properties of 1156$+$295}
\label{Linear}

\begin{figure*}
\includegraphics[height=0.9\textheight]{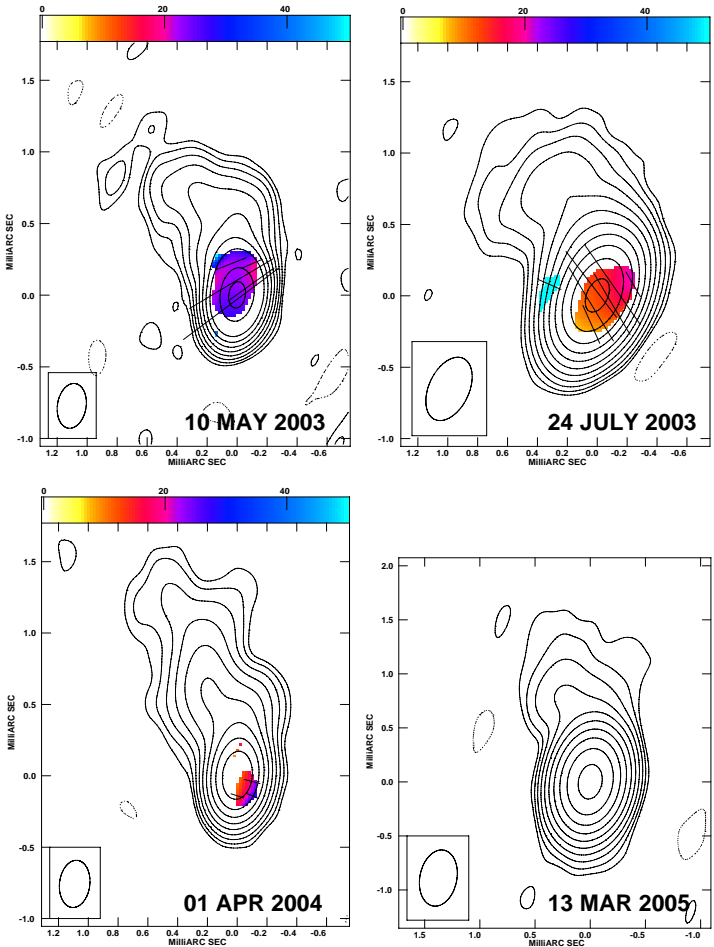}
\caption{The four epochs of VLBA images of 1156$+$295 at 43~GHz. The
contour and pseudo-color maps are similar to those in Figure
\ref{fig1}. The length of the short bars represents the strength
of polarized emission,
1~mas$\sim$4.0$\times10^{-2}$~Jy~beam$^{-1}$. The orientation of
the short bars present the apparent EVPA without rotation measure correction. The detailed parameters
of the images are listed in Tables \ref{Par15&43} and \ref{LP}.}
\label{fig2}
\end{figure*}

Figures \ref{fig1} and \ref{fig2} show the linear polarization
structure of 1156$+$295 at 15 and 43~GHz, superimposed on the
contour maps of total intensity at the corresponding frequencies.
The polarization properties of the source, the polarized flux
density $F_\mathrm{pol}$, peak polarization intensity
$F_\mathrm{peak}$, the fractional polarization $m$, and average
EVPA are given in Table \ref{LP}.

At 15~GHz, for the first three epochs (2003.36, 2003.56, and 2004.25)
there is consistent structure in polarized emission. In the inner part of
the structure, we probably see the polarized emission located in
the areas of the core C, component C6 and C7. In the core,
the fractional polarization $m$ is in the range from 2.0$\%$ to
3.6$\%$, and the average EVPA is about 110$\degr$. The jet
component C5 shows a higher fractional polarization in the range
form 6.0$\%$ to 6.7$\%$, and an average EVPA $\sim$75\degr. At
the last epoch 2005.19, the polarization structure appeared to differ
considerably from that observed during the previous
three epochs. The average EVPA is about 44$\degr$ and $m\sim4.8\%$
for the core. The change in the polarized emission structure might be
correlated with the total flux density variation as a result of the ejection
of new jet components. The enhanced polarization in C5 can
be explained as the result of a compression of the magnetic field in
this region. The compression might be caused by shocks produced by
newly ejected synchrotron plasma from the core. Under the
compression conditions, the orientations of electron vector in C5
become closer to the direction of the jet flow.

At 43~GHz, we imaged the polarization structure on a scale smaller
than 1~mas at three epochs 2003.36, 2003.56, and 2004.25. The
polarized emission near the core region at 43~GHz is shown in
Figure \ref{fig2}. At the epoch 2003.36, the polarization
properties observed at 43~GHz are consistent with those at 15~GHz,
namely EVPA$\approx120\degr$ and $m\approx2.4\%$. In the second
(2003.56) and third (2004.25) epochs, the fractional polarization is
about $2.7\%$ and $1.8\%$, respectively, but EVPA changes to
$45\degr$ and $72\degr$. The uncorrelated variation in
polarization parameters between 43 and 15~GHz close to the flare
epoch suggests that the radio activity associated with the flare
was embedded within a zone that became transparent at 43~GHz but
was still optically thick at 15~GHz at the epochs 2003.56 and
2004.25. After expansion, the active zone became optically thin
at 15~GHz at the epoch 2005.19, as indicated by the significant
changes in the polarization properties.


\begin{table}
\caption{Parameters of the hydrodynamic instability model}
\label{KHI}
\[
\begin{tabular}{cccc|cccc}
\hline \hline
        \multicolumn{4}{c|}{Assumed parameters} & \multicolumn{4}{c}{Fitted parameters} \\
        $\theta$ &$\Psi$  &$v_z$            &$y_0$ &$\alpha$  &$r_0$ &$\lambda_0$ &$\phi_0$ \\
        (\degr)  &(\degr) &(~mas~yr$^{-1}$) &(mas) &(\degr)   &(mas) &(mas)      &(\degr)\\\hline
        8.0      &3.0     & 0.19             & 0.0  & 2.6     &0.005 &0.195      &62.1\\
\hline
\end{tabular}
\]
Note: $r_0$, $\phi_0$, and $y_0$ are the initial values of the
distance along the perpendicular jet axis, the azimuth angle, and
the height along the jet axis, respectively; $\lambda_0$ is the
spatial wavelength of the initial perturbations; $v_y$ is the
velocity component along the jet axis; $\Psi$ is the half-opening
angle of the helix; $\theta$ and $\alpha$ are the viewing angle
and the position angle of the jet axis (defined as
counter-clockwise from north), respectively.

\end{table}

\begin{table}
\caption{Parameters of procession model} \label{PJB}
\[
\begin{tabular}{ccc|cccccc}
\hline \hline
\multicolumn{3}{c|}{Assumed parameters} & \multicolumn{4}{c}{Fitted parameters} \\
$\theta$ &$\Psi$  &$v_j$  & $\alpha$&P   &$t_{ref}$ &$s_{jet}$ &$s_{rot}$ \\
(\degr)  &(\degr) & ($c$) & (\degr) &(yr)&(yr)      &          & \\
\hline
8.0      &3.0     &0.98   & 2.6     &95  &56        &1         &-1\\
\hline
\end{tabular}
\]
Note: $v_j$ is the velocity of the jet component; the velocity
vector rotates around the jet axis at an angle $\Psi$ with a
period $P$; $s_{jet}$ and $s_{rot}$ are two sign parameters used
to define the jet motion direction and the helix rotation
direction; $\theta$ and $\alpha$ are the viewing angle and the
position angle of the jet axis (defined as counter-clockwise from
north), respectively.
\end{table}

\section{Helical jet structure}
\label{Hel}

Parsec-scale jets in extragalactic radio sources are often
misaligned with their kilo-parsec counterparts and show complex,
in some cases ``oscillating" structure, {\it e.g.}, 3C~273
(\citealt{Zen1988, Con1993}). A three-dimensional helical pattern
is often used to explain the phenomenology of oscillating
parsec-scale jets. In particular, bright knots might represent the
locally brighter region in the jet flow caused by enhanced Doppler
boosting effects in a helical pattern. Induced helical bending in
jets appear to be common in radio galaxies and flat-spectrum
quasars (\citealt{Con1993, Tay96, Pea1998, Lob2001}).

Theoretical interpretations of helical jets include hydrodynamic
instabilities ({\it e.g.} \citealt{Hard1986, Hard1987}) and/or
precessing jets ({\it e.g.} \citealt{Beg1980, Lin1981}).

In the hydrodynamic instability models, the helix might be triggered by small
perturbations in the accretion flow or random perturbations,
such as jet-cloud collisions at the onset of the jet. The initial
perturbations can be amplified by Kelvin-Helmholtz (K--H) hydrodynamic instability. Under
certain circumstances, the instability develops into helical modes
downstream to form the smoothly oscillating jet seen in 2-D images.
The actual evolution of the instability in the jet flow depends on
the fluctuation properties of the initial perturbations, the
dynamics of the jet flow, and the properties of the surrounding
interstellar medium (\citealt{Hard1987, Hard2003}).

In the precessing models, the helical jet trajectory is a
superposition of ballistically moving jet knots that are for example ejected
along periodically varying directions because of the
precession of the jet nozzle. This precession might have been caused by the disturbance of the jet axis by a
secondary black hole in a binary black hole system
(\citealt{Beg1980}), or from interactions between the wobbling
accretion disk and the spinning black hole (\citealt{Liu2002}).

In the following subsections, we analyze the jet structure of
1156$+$295 in the context of both possible scenarios.

\subsection{Hydrodynamic instabilities}
\label{Hyd}

A helical pattern can be produced by small
perturbations in the jet flow by a hydrodynamic instability
under certain circumstances. A dispersion relation describing the
behavior of the harmonic components of the perturbation has been
derived from a Fourier expansion of a cylindrical jet
(\citealt{Hard1987}). A particular resonant mode can grow from any
perturbations in a jet flow via K-H instability ({\it e.g.}
\citealt{zhao1992}).

To compute the oscillating jet trajectory in 1156$+$295 from
sub-pc to pc scales, we introduced a kinematic helical model on
the basis of case 2 from \citet{Ste1995}. This model corresponds
to the helical jet model derived by \citet{Hard1987}
based on the assumption of an isothermal and adiabatic expanding jet.
For a cylindrical jet, \citet{Hard1987} showed that the jet
distortions will grow in a way primarily associated with the helical
fundamental mode $n=$1 and therefore produce a helically twisted
pattern on the jet's surface. The helical jet can propagate to a
rather long distance, {\it i.e.}, 100 times the radius of the jet,
prior to the  disruption of the jet collimated flow
(\citealt{Hard1979, Hard1986, Hard1987}). Under a relativistic
approach, \citet{Ste1995} calculated the a 3-D helical trajectory
of the jet motion and assumed the bright knots in the helical jet
are enhanced by Doppler boosting, where the jet plasma is likely
to be either compressed or shocked.

Assuming that the kinetic energy, the specific momentum component along the
jet axis, and the opening half angle of the jet all remain constant, the
jet trajectory of the jet motion can be described by the equations of \cite{Ste1995} in a cylindrical coordinate system

\begin{equation}
r(t)=v_{y}t\tan\Psi+r_{0},
\end{equation}

\begin{equation}
\phi(t)=\phi_{0}+\frac{\omega_{0}r_{0}}{v_{y}\tan\Psi}\ln\frac{r(t)}{r_{0}},
\end{equation}

\begin{equation}
y(t)=v_{y}t,
\end{equation}
where $r$, $\phi$, and $y$ are the distance along the perpendicular
jet axis, the azimuth angle, and the height along the jet axis,
respectively. We note that the notation $z$, the distance along
the jet axis, in Steffen et al. (1995) has been replaced by $y$,
and is not to be confused with the cosmological redshift $z$. The
coordinates with the subscript ``$_0$'' ($r_0$, $\phi_0$, and
$y_0$) define the region where the initial perturbations take
place, $v_y$ is the velocity component along the jet axis, $\Psi$
is the half-opening angle of the helix, and $\omega_0$ is the
initial angular velocity.

The properties of the helix in this model are simply
described by six parameters: four parameters (the initial angular
velocity $\omega_0$ or, the initial spatial wavelength,
$\lambda_0=\frac{2\pi v_{y}}{\omega_{0}}$; $r_0$, $\phi_0$, and
$y_0$) describe the properties of the initial perturbations; and an additional two
parameters ($v_y$ and $\Psi$) describe the properties of the jet
in which the perturbation is propagated.

To fit the model to an observed jet trajectory, we need
two more angular parameters to describe the projection of the 3-D jet
trajectory on to the sky plane: $\theta$ the angle between the jet
axis and the line-of-sight (or the viewing angle) and, $\alpha$ the position angle of the jet axis (defined as counter-clockwise
from north).

We first assume that the perturbation takes place in a region
close to the central engine, so that $y_0$ can be set to zero for
simplifying the calculations. The viewing angles of compact jet
components calculated in Section 3.4 vary between $1.5\degr$ and
$22.1\degr$. If we include published data at lower observing
frequencies in \citet{Hon2004}, the arithmetic mean value of the
viewing angles of compact jet components is $8.0\degr(\pm 4.9)$.
We take the mean value as an appropriate estimate of the viewing angle of the
helix axis $\theta$. The half-opening angle $\Psi$ of 1156$+$295
is estimated to be $\sim3\degr$ (determined from the maxima range
of the position angles and viewing angles of bright jet
components).


Furthermore, we derived the jet velocity $v_j$ in Section
\ref{Phy}. Since both the opening angle of the helix cone and the
viewing angle of the jet are small, the velocity along the jet
axis is approximately equal to the jet velocity, $v_y\approx v_j
\sim 0.98~c$. Taking into account of the effect in special
relativity, the apparent jet velocity is boosted by a factor of
$\displaystyle \frac{\sin\theta}{1-v_{y}\cos\theta/c}\simeq4.7$,
{\it i.e.}, ${\displaystyle
v_y^{ap}=v_y\frac{\sin\theta}{1-v_{y}\cos\theta/c}\simeq4.6~c}$,
or a proper motion of 0.19~mas~yr$^{-1}$.

The remaining four parameters ($\alpha$, $r_0$, $\phi_0$, $\lambda$) are the free
parameters to be determined by fitting the model to the observed
jet trajectory. As illustrated in Figure \ref{fig7}, the observed
jet trajectory can be well fitted with the isothermal helical jet
model on wide scales ranging from 0.08~mas (0.6~pc) to 35~mas
(250~pc). The symbols represent the jet components detected from
the 15 and 43~GHz VLBI data from 2003 to 2005 presented in this paper, as well as
the ones detected at 1.6, 22, and 86~GHz from other papers (\citealt{Hon2004, Jor2001,
Lee2008}).

The best-fit model parameters are given in Table \ref{KHI}. The
position angle $\alpha\approx2.6\degr$ in the source frame is
inferred from our model fitting, and is consistent with the
observed jet structure that extends northward from the
core.

Our model fitting suggests that there is a region with
$r_0\approx0.005$~mas (or $\sim$0.04~pc) in which the jet flow
might be perturbed substantially. This compact region in the
vicinity of the central engine might be close to or associated with
the jet nozzle. This radius (0.04~pc) is about 1000 times of the
Schwarzschild radius ($R_s\approx4\times10^{-5}$~pc) of a supermassive black hole (SMBH) of
4.3$\times10^8 M_{\odot}$ which has been inferred to exist in 1156$+$295
(\citealt{Pia2005}). It appears to be much greater than the inner edge of
the accretion disk radius of 3 $R_s$, equivalent
to the radius of the inner most
stable orbit, but comparable to the typical size of the BLR.
Perturbations in the jet flow might be imposed by the interactions
between the jet plasma and BLR clouds.

We also examined the other cases considered by \citet{Ste1995}. In
case 1, the specific angular momentum, the Lorentz factor, and the
momentum component along the jet axis are assumed to be conserved.
This case is rejected at first since no more than one quarter
revolution can be produced for the helical pattern. Both cases 3
and 4 involve a magnetic field. case 4 takes into account the
dissipation of the kinetic energy of the jet flow by means of
collisions with the surrounding ISM and radiative losses, while case 3
considers the variation in the specific momentum. However, we
found that the fitted curves in the models of both cases 3 and 4 degenerate
quickly into a straight line on the scale of several milliarcsecs,
inconsistent with the multiple jet bendings observed in 1156$+$295
on the scales ranging from mas to arcsecond.

\subsection{Precessing jet}
\label{Pre}

Alternatively, the observed helical jet morphology might represent a
superposition of ballistic jet components ejected sequentially
from a precessing jet nozzle ({\it e.g.} \citealt{stir2003}). On
the basis of the assumption that the jet components are ejected from a
precessing jet nozzle and move along ballistic paths, we fit the
bright knots observed in the jet with a simple precessing model
(\citealt{Hje1981, Bar1983, An2010}).

In this model, the jet knot moves outward from the central object
with a constant velocity $v_j$. The velocity vector $v_j$ rotates
around the jet axis at an angle $\Psi$ and a period $P$. Each jet component travels along a ballistic trajectory, and the
projection of the superposition of their ballistic trajectories
constitutes the observed helical jet trajectory on the sky plane.
The angle between the jet axis and the line of sight is $\theta$
and the observed jet axis lies at a position angle $\alpha$.

The velocity of a given jet can be described as in \cite{Hje1981} with

\begin{equation}
v_x=s_{jet}
v[\sin\Psi\cos\Omega(t-t_{ref})\cos\theta+\cos\Psi\sin\theta],
\end{equation}

\begin{equation}
v_y=s_{jet} v\sin\Psi\sin\Omega(t-t_{ref}),
\end{equation}

\begin{equation}
v_z=s_{jet}
v[\cos\Psi\cos\theta-\sin\Psi\cos\Omega(t-t_{ref})\sin\theta],
\end{equation}
where $s_{jet}$ is used to define the jet motion direction and
$s_{jet} =+1$ corresponds to moving towards the observer,
$t_{ref}$ is the reference time, $\Omega$ is the rotating
angular velocity of the jet velocity vector,
$\Omega=s_{rot}2\pi/P$, $s_{rot}$ is used to define helix rotation
direction, and $s_{rot}=+1$ corresponds to counterclockwise.

The transverse velocity of the jet can be decomposed into two
orthogonal components of right ascension ($\alpha$) and
declination($\delta$)

\begin{equation}
v_{\alpha}=\sin\alpha v_y+ \cos\alpha v_z,
\end{equation}

\begin{equation}
v_{\delta}=\cos\alpha v_y- \sin\alpha v_z.
\end{equation}

Taking the relativistic aberration into account, the proper motion of a given component ejected at time $t_{eject}$ on the sky
plane can be written as:

\begin{equation}
\mu_{\alpha}=\frac{v_\alpha(t-t_{eject})/\cos\delta}{d(1-v_x/c)},
\end{equation}

\begin{equation}
\mu_{\delta}=\frac{v_\delta (t-t_{eject})}{d(1-v_x/c)},
\end{equation}
where $d$ is the distance from the observer to the source.

A jet velocity $v_j \geq 0.98c$ is derived in Section 3.4 in the
source-rest frame. The jet half-opening angle $\Psi$ of $3\degr$
and viewing angle of $\theta=8\degr$ are the same as the values
adopted in our K-H instability model (Section \ref{Hyd}). Since the jet is moving toward us, $s_{jet}$ is set as
$+1$. Adopting an eyeball approach, we fit the precessing model to the jet structure
observed on the parsec scale by adjusting the remaining free
parameters (position angle of the helix $\alpha$, precessing
period $P$, reference time $t_{ref}$, and the sign of rotation
$s_{rot}$).

The fitted parameters are given in Table \ref{PJB}. As illustrated
in Figure \ref{fig8}, on the scale of tens of~mas, the precessing jet
model provides a reasonable fit to the observed data but the inner
2-mas jet structure appears to be poorly fitted. The structure on
the scale $>35$~mas is also difficult to fit with the
precessing jet model. The kinematic model described here represents
an ideal case, in which the jet is continuously ejected at a
constant velocity and the jet ejection direction varies
periodically. However, the actual jet precession in a powerful
radio source might be more complex. For example, the period of the
jet ejecting direction may be altered by intermittent radio
outbursts and jet ejections (\citealt{Rey1997}). The jet may
interact with the BLR and/or NLR clouds to generate stationary
shocks. Abrupt changes in the jet direction and violent jet-ISM
interactions have been observed in some compact-steep-spectrum
sources such as 3C~48 (\citealt{An2010}) and 4C~41.17
(\citealt{Gur1997}). In addition, the intrinsically small jet
bending in the inner 2~mas might be enhanced by projection
effects. There is also a possibility that hydrodynamical
instabilities of the plasma at the onset of the jet may modulate
the fine structure of the whole curved jet.

A precessing jet from AGNs is a characteristic phenomenon in a
binary black hole (BBH) system (\citealt{Beg1980}). However, the
orbital period of the BBHs remains uncertain because the relationship
between the precessing period and orbital period is unclear. Using
the ratio of the precession period to the orbital period in X-ray
binaries as an analogy (\citealt{Mar1980}), the precessing period
fitted for 1156$+$295 gives a upper limit to the orbital period of
the BBHs of about 95 years.

Assuming Keplerian motion, the derived semi-major axis, or the
largest separation between the two black holes, is $<$0.08 pc for
a primary black hole with a mass of 4.3$\times10^8 M_{\odot}$.
This size scale is one tenth of the current VLBI resolution.
Moreover, the real orbital period is probably a few percent of the
precessing period, thus the orbital separation might be a
few $\times$0.01 pc.

Some theoretical calculations ({\it e.g.} \citealt{Yu2002}) show
that the orbital parameters of a survival BBH depend on the
galactic velocity dispersions and blackhole mass ratios. An
orbital period of a few tens of years suggests a mass ratio of the
secondary to primary black holes of between 0.01 and 0.05. Combining
the close separation of the binaries and the relatively low mass
of the secondary BH, the radio emission from the secondary BH
might be very weak and below our current detectable limit in VLBI.
Evidence of possible BBHs in 1156$+$295 may be represented by possible periodic variabilities in the radio, optical,
and X-ray light curves, and the expected apparent variability timescale would
be suppressed by relativistic aberration resulting in an orbital period much
shorter than the intrinsic value (\citealt{Yu2002}).

\subsection{Comparison of the two models}
\label{Comp} In the hydrodynamic instability model, the jet
trajectory on scales from sub-mas to several tens of mas and
even larger can be naturally explained as the pattern produced by
the fundamental mode ($n=1$) or a helical mode in K-H instability
model with an initial characteristic wavelength of
$\lambda_0=0.2$~mas (1.5~pc). On the basis of our model fitting to
the data, the helical mode appears to be excited at the jet base
on a scale of 0.005~mas or 10$^3R_s$, within the scope of the BLR.
The emission from the jet knots moving towards the direction of
observers is Doppler boosted. In comparison, the precessing jet
model alone appears to have difficulty in fitting the jet
trajectory on the observed scales ranging from sub-mas, several
tens of mas, up to arcsec. In particular, the multiple bends
observed on the mas scale cannot be explained solely with the
precessing jet model. With additional mechanisms such as
interactions of jet plasma with the interstellar medium, we may be
able to explain the considerable deviation of the observed
jet trajectory from that predicted in the precession jet
model. Furthermore, the jet trajectory observed beyond 35~mas
imposes a further limit on the precessing jet model using a single
rotating angular velocity ($\Omega$) of the precessing jet nozzle.
To fit the  precessing jet model to the data observed beyond
35~mas suggests that an additional rotating period
(${\displaystyle P=\frac{2\pi}{\Omega}}$), which is much longer
than 95~years derived from the inner jet, is required. For 1156$+$295,
despite the deceleration, the jet plasma on pc scales beyond
35~mas has been ejected from the jet nozzle for at most a few hundred years. It would be
incredible if the period of a process associated with the
nuclear dynamics had changed considerably and abruptly in such a short period.
Finally, based on our VLA observations
(\citealt{Hon2004}), the jet trajectory becomes a straight line on
the kpc scale (or $>$0.1 arcsec). In contrast, the precessing jet
model predicts that the curvature of the jet trajectory projected
on the sky becomes large unless  ballistic jet components ejected
from the precessing nozzle in a wide range of directions had been
lined up, on the way travelling to the outer region on the kpc
scale, by a tunnel in the ISM or by a poloidal magnetic field in
1156+295.

In short, although the results from our model fitting support the idea that
the K-H instability is responsible for the observed helical jet
pattern, we cannot fully rule out the possibility that the
oscillating jet structure observed between 1~mas (10~pc) and 40~mas
(300~pc) is driven by a precessing jet nozzle with a rotating
period of 95 yr. Additional astrophysical processes ({\it e.g.} interaction between the
jet plasma with the the BLR and/or NLR clouds) should be taken
into account to explain the complex bending jet within 10~pc, and
recollimation of the jet beyond 300 pc, if the jet precessing
model is to be successfully applied to 1156$+$295.


\section{Summary}
\label{Sum} We observed 1156$+$295 with the VLBA at the
wavelengths of 86, 43, and 15~GHz with angular resolutions up to
0.08~mas (0.5~pc) at four discrete epochs during 2003 to 2005.
Based on the analysis of the new VLBA data from the multiple-epoch
and multiple-frequency observations, we have presented new results on
the morphology, physical properties, and kinematics of the jet
observed on the scales from sub-pc to pc from the blazar. We have also
discussed the astrophysical implications for the helical pattern
of the jet emission between sub-pc and kpc by fitting
both the K-H instability model and the jet-nozzle precessing model
to the observed data.

The core-jet structure was detected in 1156$+$295 at 43 and 15~GHz
at angular resolutions from 0.2~mas to 0.6~mas. Six jet components
were identified. Among the six, the innermost three were detected
for the first time. The apparent transverse velocities of the six
jet components range between 3.6~$c$ and 11.6~$c$. At 86~GHz, during the
first epoch the source was resolved into a core-jet structure on the image,
during the last epoch the source seems to be resolved into a double-point source
in the $uv$-domain, and during other epochs only a point-like source was detected
on both the image and the $uv$-domain. Ultra-high apparent brightness
temperatures of $>10^{12}K$ were inferred suggesting that highly
relativistic jet plasma ejected from the nucleus moves in the
directions close to the line-of-sight of observers. After correcting for
both the Doppler beaming and cosmological redshift effects, the
intrinsic brightness temperature of the nuclear radio emission is
below the inverse Compton catastrophe limit. During the observing
period from 2003 to 2005, a radio flare was present in the data,
suggesting that a radio activity of the blazar originated inside.
Linear polarization was detected at 43 and 15~GHz in both the core
and inner jet. During the quiescent epochs, the EVPA of the jet
polarized emission differed by $\sim35\degr$ from that of the core.
No significant change in EVPA of the polarized emission from the
core between 2~cm and 7~mm was observed during the quiescent
epochs. At the epochs after the radio flare, EVPA observed at 7~mm
from the core shows a significant change from that of the
quiescent epoch.

By model fitting a helical fundamental mode of K-H instability
and a precessing jet-nozzle, we have found that the
K-H instability model successfully fits the observed jet
trajectory on the scale ranging from sub-pc to kpc. Our
results suggest that the fundamental helical mode with an initial
characteristic wavelength of 0.2~pc appears to be excited at the
jet base on a scale of 0.005~pc or $10^3R_s$, a typical BLR size for
a SMBH of 4.3$\times10^8$ M$_\odot$. The precessing jet
model only fits the jet trajectory observed on a scale ranging
from 10~pc to 300~pc and requires additional astrophysical
processes to explain the complex jet bending observed in the
inner region (sub-pc to 10~pc) and the re-collimation of the
large scale jet ($>300$ pc).


\begin{acknowledgements}
WZ, XYH, and TA are grateful for the support from the National
Natural Science Foundation of PR China (NSFC10473018, 10503008),
the Ministry of Science and Technology of China (Grant
No.2009CB824900/2009CB24903) and the eScience program "Fast eVLBI
Imaging and dUT1 Determination" of the Chinese Academy of
Sciences. WZ thanks Giuseppe Cim\`{o} for the help of calibration
for the polarization VLBA data. WZ thanks JIVE for the hospitality
during her visit to the Netherlands for the data reduction in
2007. The VLBA and VLA are instruments of the National Radio
Astronomy Observatory, a facility of the US National Science
Foundation operated under cooperative agreement by Associated
Universities, Inc. This research has made use of data from the
MOJAVE database that is maintained by the MOJAVE team (Lister et
al., 2009, AJ, 137, 3718). This research uses the data from
the University of Michigan Radio Astronomy Observatory which has
been supported by the University of Michigan and by a series of
grants from the National Science Foundation, most recently
AST-0607523. The Owens Valley Radio Observatory Millimeter Array
is supported in part of the National Science Foundation NSF Grant
AST99-8154. Thanks are given to the anonymous referee for useful comments and
suggestions, and Claire Halliday (language editor of
A\&A) for proof-reading of the text.

\end{acknowledgements}

\clearpage

\begin{figure*}
\label{fig4}
\includegraphics[width=0.9\textwidth]{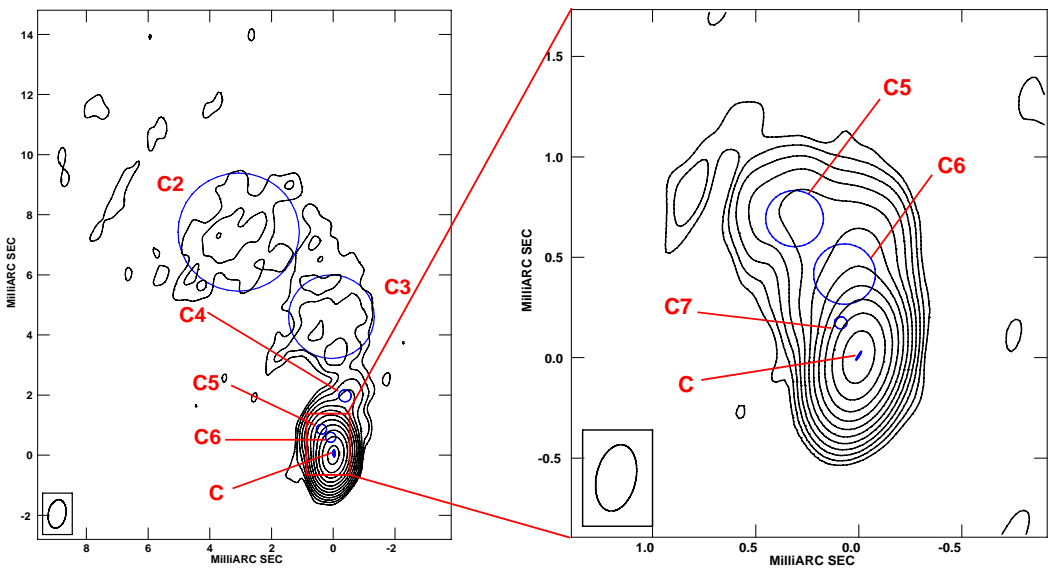}
\caption{The fitted models are overlaid on the clean images of 15
and 43~GHz on epoch 2003.36 to indicate their positions and intrinsic size.}
\label{fig4}
\end{figure*}

\begin{figure*}
\includegraphics[angle=0, width=1.0\textwidth]{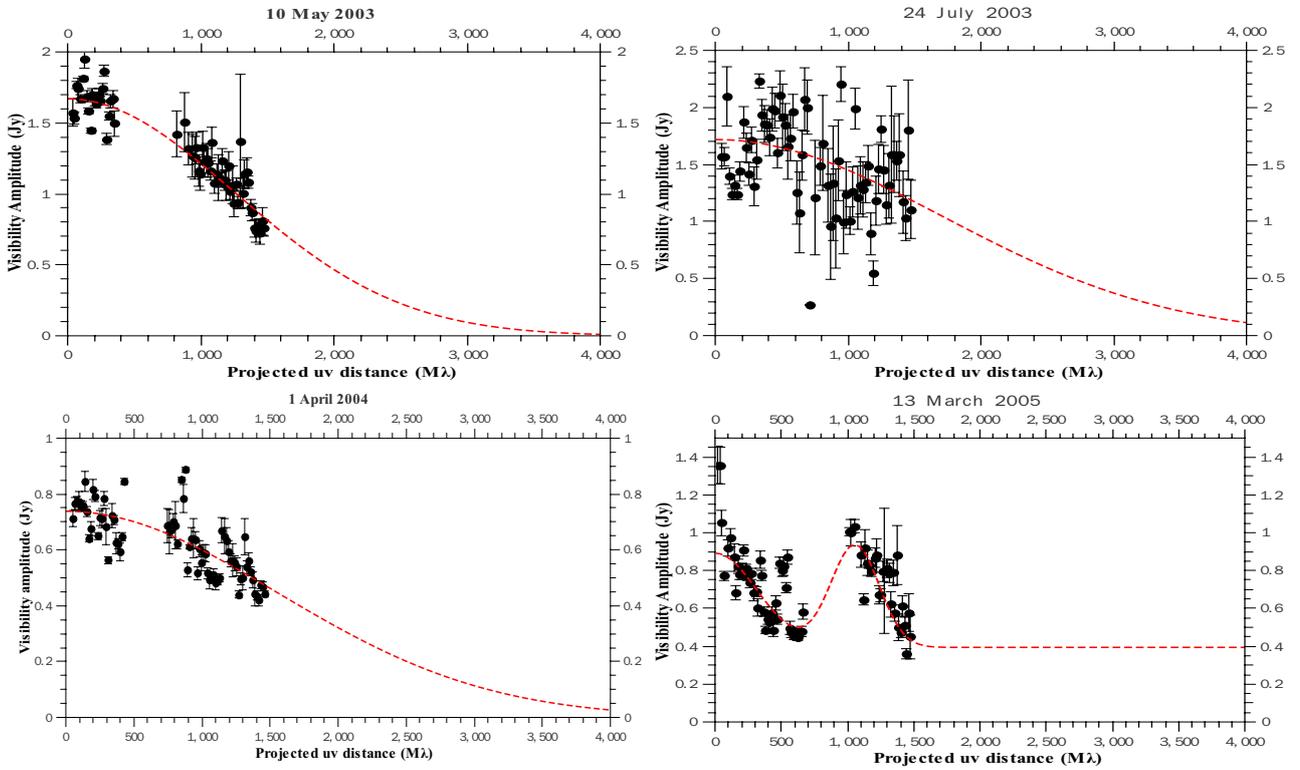}
\caption{Estimating the flux density ($S_{f}$) and size
($\theta_{d}$) of the source using the brightness distribution
models introduced by \citealt{Pea1995}): for the first three
epochs, an extended source model could be used, whereas for the last epoch,
a double-point source model with an intensity ratio of 3:1 and an angular separation of 0.16~mas were used}
\label{86GHzNon}
\end{figure*}

\clearpage
\begin{figure}
\includegraphics[angle=0,width=0.45\textwidth]{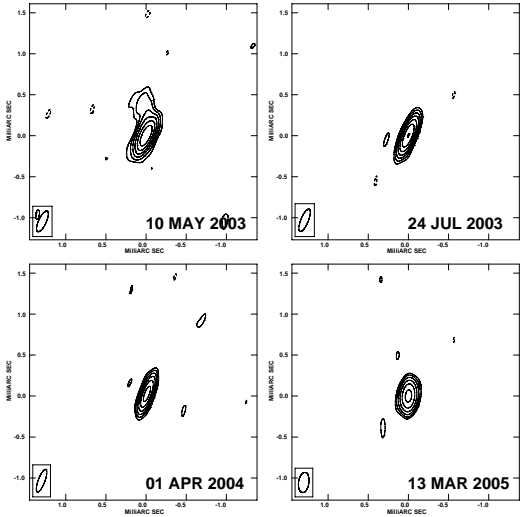}
\caption{The four epochs VLBA images of 1156$+$295 at 86~GHz. The
detailed parameters of the images are listed in Table
\ref{Par86}.} \label{fig3}
\end{figure}

\begin{figure}
\includegraphics[angle=0, width=0.45\textwidth]{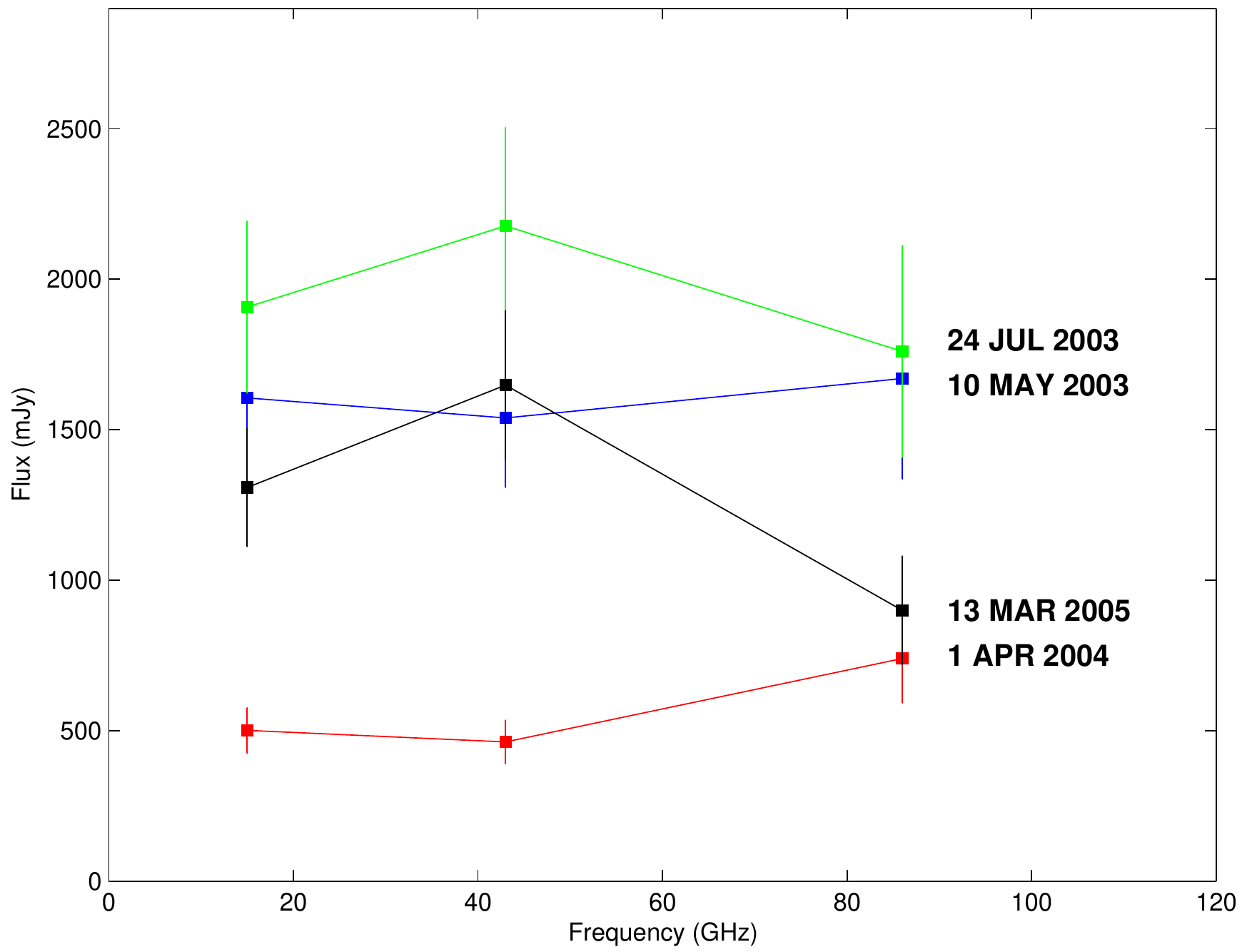}
\caption{A spectral plot of the core component at the four epochs.
The error bars show errors in the measurement of the flux density of
the core component. } \label{fig4b}
\end{figure}

\begin{figure}
\includegraphics[width=0.45\textwidth]{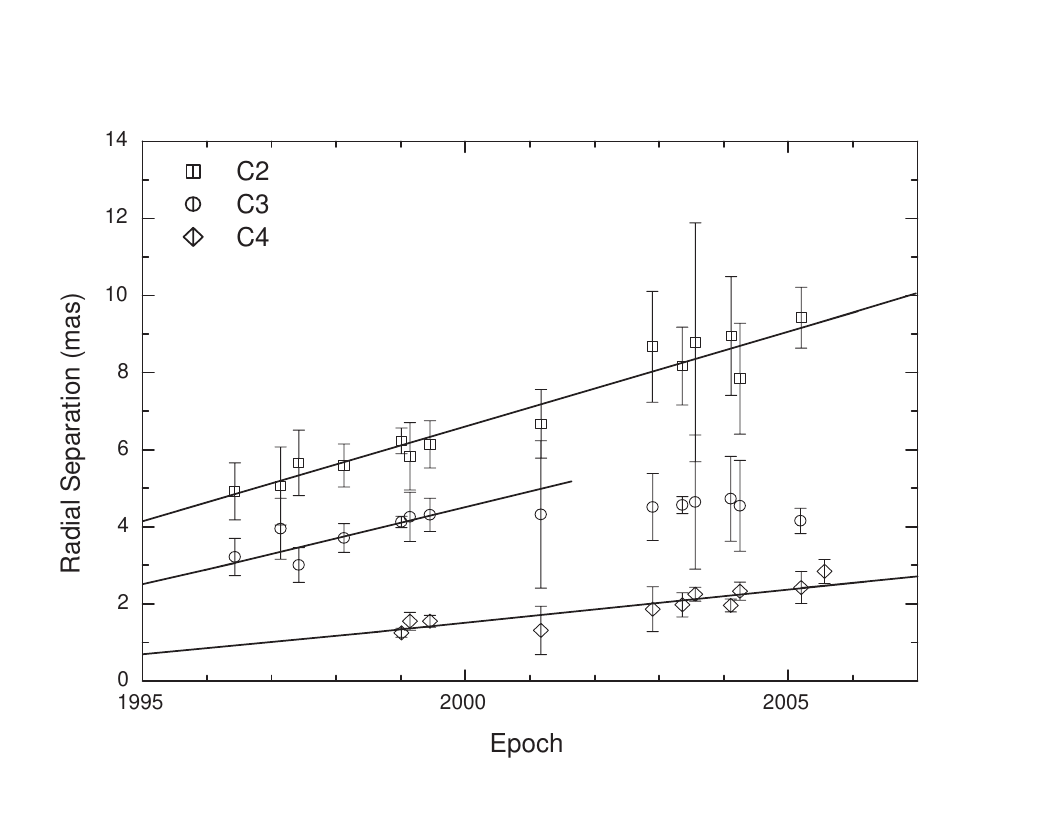}
\caption{The proper motions of the outer jet components C2, C3, and C4
are 0.49$\pm$0.03 mas~yr$^{-1}$, 0.41$\pm$0.08 mas~yr$^{-1}$, and
0.18$\pm$0.03 mas~yr$^{-1}$, which correspond to 11.6$\pm$0.7~$c$,
9.7$\pm$1.9~$c$, and 4.3$\pm$0.7~$c$, respectively.} \label{fig51}
\end{figure}

\begin{figure}
\includegraphics[width=0.45\textwidth]{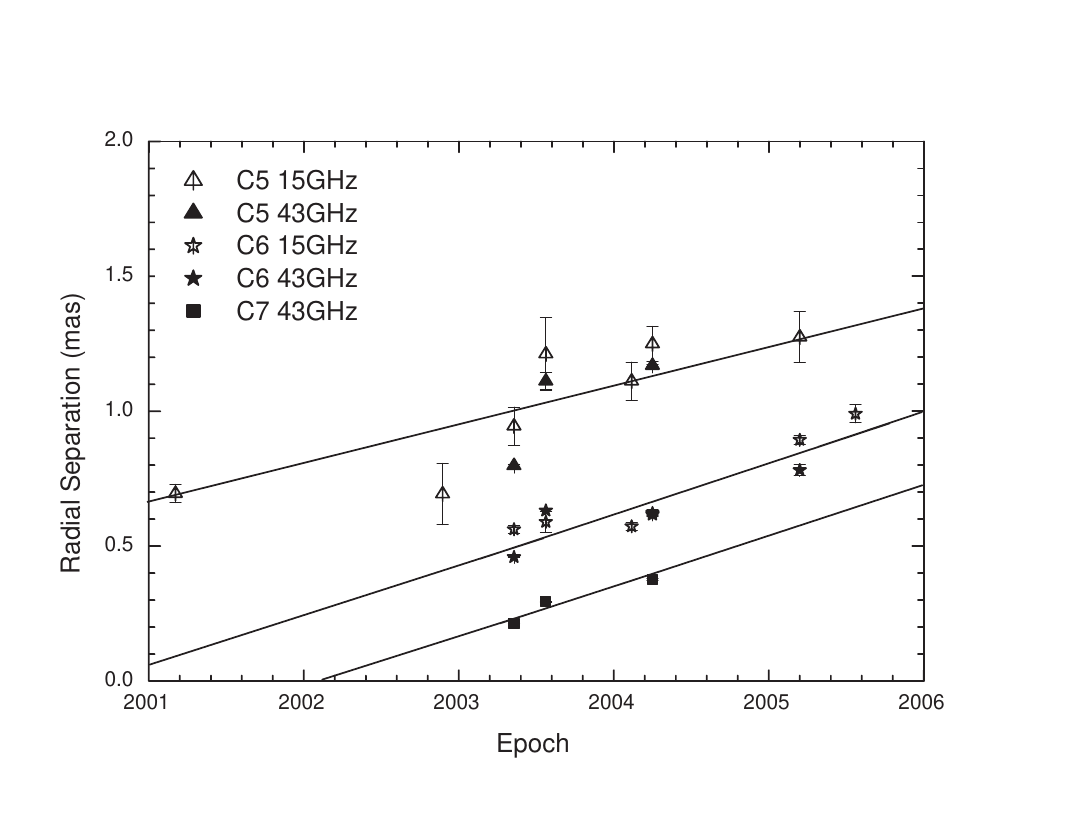}
\caption{The proper motions of inner jet components C5, C6, and C7
are 0.15$\pm$ 0.02 mas~yr$^{-1}$, 0.19$\pm$0.02 mas~yr$^{-1}$, and
0.16$\pm$0.04 mas~yr$^{-1}$, which correspond to 3.6$\pm$0.5~$c$,
4.5$\pm$0.5~$c$, and 3.8$\pm$0.9~$c$, respectively.} \label{fig61}
\end{figure}

\begin{figure*}
\includegraphics[angle=0,width=0.8\textheight]{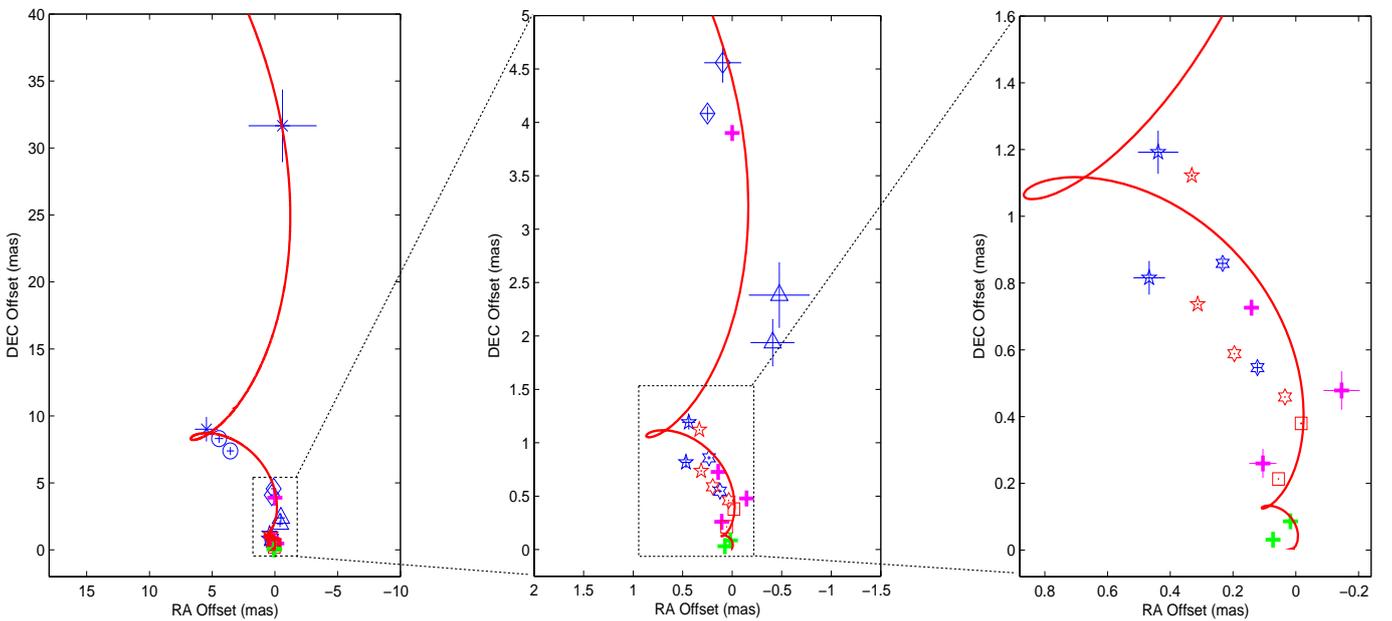}
\caption{The best fitting of a model with the fundamental helical
mode in K-H instability (red curve) to the data (symbols). The
color of symbols represents the data taken at different
frequencies: blue for 15~GHz or lower frequencies, magenta for
22~GHz, red for 43~GHz, and green for 86~GHz. In addition,
different types of symbol mark different jet components: star for
C0, cross for C1, open circle for C2, open diamond for C3, open
triangle for C4, open pentagram for C5, open hexagram for C6, and
open square for C7. The components (B3, B2, C, D1) from Jorstad's
paper and (A1, A2) from Lee's paper are marked with plus signs
with red and pink, respectively.}\label{fig7}
\end{figure*}

\begin{figure*}
\includegraphics[angle=0,width=0.8\textheight]{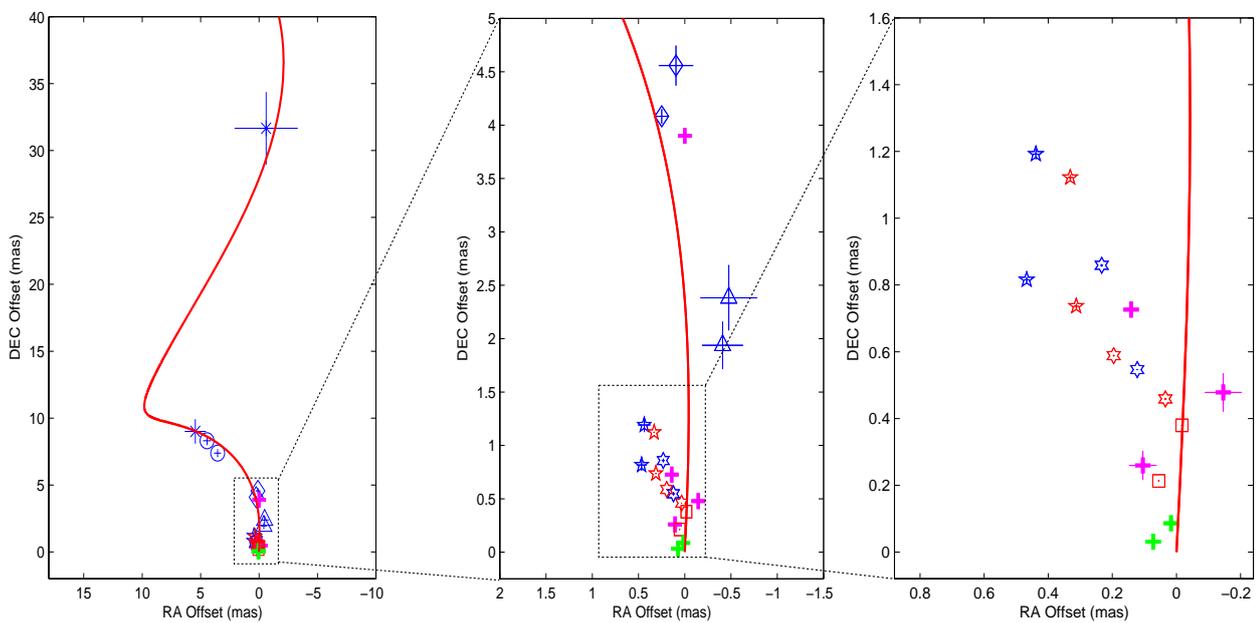}
\caption{The best fitting of the precessing-jet model (red curve)
to the same observed data as described in Figure \ref{fig7}. The
color of symbols represents the data taken at different
frequencies: blue for 15~GHz or lower frequencies, magenta for
22~GHz, red for 43~GHz, and green for 86~GHz. In addition,
different types of symbol mark different jet components: star for
C0, cross for C1, open circle for C2, open diamond for C3, open
triangle for C4, open pentagram for C5, open hexagram for C6, and
open square for C7. The components (B3, B2, C, D1) from Jorstad's
paper and (A1, A2) from Lee's paper are marked with plus signs
with red and pink, respectively. } \label{fig8}
\end{figure*}

\end{document}